\documentclass[twocolumn,pre,superscriptaddress,showpacs,preprintnumbers,amsmath,amssymb]{revtex4}

\usepackage{graphicx}
\usepackage{dcolumn}
\usepackage{bm}
\newcommand{\mct}{_{c}}
\newcommand{\simmct}{_c^{\mbox{\scriptsize (sim)}}}
\newcommand{\themct}{_c^{\mbox{\scriptsize (theory)}}}

\newcommand{\e}{\mbox{\large e}}
\newcommand{\kb}{k_{\mbox{\scriptsize B}}}

\newcommand{\ave}[1]{\langle {#1} \rangle}

\begin{document}
\title{Slow Dynamics of the High Density Gaussian Core Model}
\author{Atsushi Ikeda}
\affiliation{Institute of Physics, University of Tsukuba, Tennodai 1-1-1, Tsukuba 305-8571, Japan}
\author{Kunimasa Miyazaki}
\affiliation{Institute of Physics, University of Tsukuba, Tennodai 1-1-1, Tsukuba 305-8571, Japan}

\date{\today}
\begin{abstract}
We numerically study crystal nucleation and glassy slow dynamics 
of the one-component Gaussian core model (GCM) at high densities.
The nucleation rate at a fixed supersaturation
is found to decrease as the density increases. 
At very high densities, the nucleation is not observed at all in the 
time window accessed by long molecular dynamics (MD) simulation. 
Concomitantly,  the system exhibits typical slow dynamics of the
supercooled fluids near the glass transition point. 
We compare the simulation results of the supercooled GCM with the
predictions of mode-coupling theory (MCT) and find that the agreement
between them is  better than any other model glassformers studied
 numerically in the past.  
Furthermore, we find that a violation of the Stokes-Einstein relation 
is weaker and the non-Gaussian parameter is smaller than canonical glassformers.  
Analysis of the probability distribution of the particle displacement 
clearly reveals that the hopping effect is strongly suppressed in the high
 density GCM. 
We conclude from these observations that the GCM is more amenable to the mean-field
 picture of the glass transition than other models. 
This is attributed to the long-ranged nature of the interaction
 potential of the GCM in the high density regime. 
Finally, the intermediate scattering function at
small wavevectors is found to decay much faster than its self part, 
indicating that dynamics of the large-scale density fluctuations 
decouples with the shorter-ranged caging motion.
\end{abstract}
 \maketitle

\section{Introduction}

Essential aspects of the glass transition of the supercooled liquids remain still
elusive despite of decades of study.
Many theories and scenarios have been proposed to explain 
the dramatic slow down of the systems and the 
associated growing cooperative length scales near the glass transition
point~\cite{Debenedetti2001,Cavagna2009b,Biroli2009,Berthier2011d}. 
They can explain the experimental results 
equally well or equally poorly but none of them have been proved to be decisively better than other. 
Even a satisfactory mean-field picture of the glass transition 
has not been established~\cite{Ikeda2010,Schmid2010b}.   
Numerical simulation of simple model fluids is an ideal route to examine 
the competing theories.
Considerable efforts have been put forward to gain insight from the
dynamical behaviors of simple model glassformers {\it in silico}, but compelling answers are
still lacking. 
There are several reasons why the simulation studies are 
not successful in sorting out numerous scenarios and theories. 
First, the model systems are more or less similar; 
the pair potentials of canonical glassformers studied in the past are exclusively 
characterized by short-ranged strong repulsions. 
Examples are Lennard-Jones, its WCA counterpart, soft-core, and the
hard sphere potentials. 
Since the strong repulsion dominates thermodynamic and dynamic properties of
dense fluids, it is hardly surprising that the results for these models
are 
qualitatively similar~\cite{Andersen2005,Berthier2009d}.
Studies of a completely different class of potential systems may potentially 
diversify our views and perspectives on the glass transition within
the limited accessible time windows of the simulations.
Secondly, the model systems are not clean enough. 
Even the simplest class of model glassformers (with a few
exceptions~\cite{Sausset2010b,Charbonneau2010}) 
are inevitably bidisperse or polydisperse 
in order to avert the
nucleation to the crystalline phase~\cite{Andersen2005}. 
This complicates quantitative assessment of the simulation results. 
Finally, we still lack a realistic model glassformer which conforms 
to the mean-field picture in finite dimensions. 
Concept of the mean-field scenario of the structural glass transition is
basically borrowed from the mean-field theory developed in the spin
glass
communities~\cite{Kirkpatrick1989,Cavagna2009,Biroli2009,Berthier2011d}.   
The replica theory~\cite{Mezard1999,Parisi2010} and mode-coupling
theory (MCT)~\cite{Gotze2009} are believed to be the static and dynamic
versions of the mean-field theory of the glass transition, simply
because of their apparent resemblance to the spin-glass counterparts. 
The mosaic pictures of the random first order transition theory 
has been developed as the finite dimension version of this mean field
pictures~\cite{Kirkpatrick1989,Lubchenko2007,Biroli2009}.  
Accumulated simulation data are not inconsistent 
qualitatively from the prediction of the mean field theories but
the quantitative agreement between simulation results and theoretical predictions 
are far from compelling. 
The best way to verify the mean-field scenario would be to 
take the mean-field limit by either going to higher dimensions or making 
the system's interactions longer-ranged. 
Recently, simulations for four dimensional systems have been
performed~\cite{Eaves2009,Charbonneau2010}. 
Results therein hint that the dynamic heterogeneities are suppressed
compared with three dimensional systems
and agreement with MCT moderately improves~\cite{Charbonneau2010}. 
However, considering the current computational abilities, 
it would be hard to simulate the system beyond four dimension, whereas
the upper critical dimension of the glass transition is argued to be
eight~\cite{Biroli2007b,Biroli2006b}. 
On the other hand, few studies have been done for realistic liquids with
long-ranged particle interactions~\cite{Zaccarelli2008b,Dotsenko2004,Mari2011}. 

The Gaussian core model (GCM) is a candidate to dispel all of the
above-mentioned concerns and could be an ideal and clean bench to
compare with various glass theories. 
The GCM consists of the point particles interacting with a Gaussian shaped
repulsive
potential~\cite{Stillinger1976,Stillinger1997,Lang2000,Louis2000b,Prestipino2005,Mladek2006,Mausbach2006,Zachary2008,Krekelberg2009c,Shall2010}; 
\begin{eqnarray}
v(r) = \epsilon \exp[-(r/\sigma)^2], 
\end{eqnarray}
where $r$ is the interparticle separation, 
$\epsilon$ and $\sigma$ are the parameters which characterize the energy
and length scales, respectively. 
The GCM is one of the simplest models of the so-called ultrasoft potential
systems which are characterized by the bounded and long-tailed repulsive
potential~\cite{Likos2001}.
Recently, we have reported that the one-component GCM vitrifies at very
high densities~\cite{Ikeda2011}.
The GCM or the ultrasoft particles in general 
have very distinct and exotic properties both thermodynamically and
dynamically~\cite{Stillinger1976,Stillinger1997,Lang2000,Louis2000b,Prestipino2005,Mladek2006,Mausbach2006,Zachary2008,Krekelberg2009c,Shall2010,Ikeda2011,Ikeda_I},   
such as the re-entrant melting at high densities, negative thermal
expansion coefficient, and anomalous density dependence of the 
diffusion coefficient. 
There are several studies on the glass transition of the ultrasoft
particles~\cite{Foffi2003b,Zaccarelli2005c,Berthier2009c,Berthier2010i}
and it was found that they exhibit rich dynamical behaviors
different from conventional model glassformers~\cite{Foffi2003b,Zaccarelli2005c}. 
One of the advantages to study the glass transition of the ultrasoft particles is that, 
due to the mild repulsion tail of the potential, the density as well as
the temperature can be used as a parameter to control the system. 
Exploring the wide range of density--temperature parameter space
makes it easier to establish various scaling laws, to bridge
the gaps between temperature-driven ordinary glasses and density-driven colloidal glasses, 
and to help unifying the concepts of the finite-temperature glass
transition and the zero-temperature jamming transition~\cite{Berthier2009c,Berthier2010i}. 
However, most studies in the past focused on the relatively low density
regime, where the generic nature of the glass transition is not
extremely different from that of the conventional model glassformers. 
The systems at low densities, including the GCM, also had to be either
polydisperse or bidisperse in order to avoid crystallization. 

The GCM at very high densities is very different~\cite{Ikeda2011}. 
First of all, the system vitrifies without poly(bi)dispersity.
The nucleation rate systematically decreases as the density increases
and the system starts exhibiting typical slow dynamics observed in
supercooled fluids near the glass transition point.
Furthermore, the dynamics is quantitatively well-described by
MCT. 
Especially, the MCT nonergodic transition point extracted from the
simulation unprecedentedly matches with the theoretical prediction. 
Besides, the violation of the Stokes-Einstein relation and the amplitude
of the non-Gaussian parameter, both of which is the manifestation of the 
heterogeneous fluctuations of dynamics, are suppressed. 
We conjecture that these facts can be attributed to the long-ranged
nature of the interaction potential at the high densities where
particles are overlapped.  
These results suggest that the high density GCM is not only one of the
cleanest model glassformers {\it in silico}, but also the closest to the
mean-field model.  

In this paper, we present thorough and complete numerical analysis of 
the nucleation and glassy dynamics of the high-density and one-component GCM. 
We not only present the exhaustive set of the
numerical results but also provide with the new evidence which 
bolsters the validity of MCT.
Detailed analysis of thermodynamic and structural properties of the high
density GCM, such as the phase diagram and the static structure factors 
are discussed in Ref.~\cite{Ikeda_I}. 
In the previous study~\cite{Ikeda2011}, we have attributed the 
weak violation of the SE relation and smaller non-Gaussian parameter to
the suppression of the dynamic heterogeneities.
We provide stronger and more direct evidence that intermittent heterogeneous
motion is suppressed by monitoring the distribution of the particle
displacement as a function of time.
We also evaluate the correlation functions of single and
collective density fluctuations. 
Surprisingly we find that dynamics of the collective density
decouple from the single particle density at large length scales, where 
the former relaxes much faster than the latter.
This is in stark contrast with the ordinary model glassformers
for which the slow glassy dynamics set in over the whole length
scales for both collective and single particle densities alike.
We compare these simulation results with MCT predictions and 
find that MCT beautifully captures the decoupling of dynamics at the large
length scales. 
However, we also find a subtle but noticeable disagreement 
of MCT from the simulation results at intermediate length scales, where 
the nonergodic parameter (the plateau height of the 
two step relaxation in the density correlators) 
predicted by MCT shows a weak shoulder which tends to grow as the density increases. 
This shoulder is reminiscent of those found for the $d$-dimensional hard sphere glasses 
at large $d$ evaluated from MCT~\cite{Ikeda2010,Schmid2010b} and may be
a signal of breakdown of MCT at the mean field limit. 

This paper is organized as follows. 
In Sec.~II, we summarize the simulation method, theoretical background,
and the setting of the system.
The nucleation dynamics from fluid to crystalline phase is discussed in
Sec.~III.  
In Sec.~IV, we present all simulation results on various static and
dynamical observables. 
Detailed analysis and careful comparison of the simulation results with
the MCT predictions are made.
Suppression of the dynamic heterogeneities are also discussed. 
Finally, Sec.~IV concludes the paper with a summary. 

\section{Preliminaries}

\subsection{Simulation Methods}

\begin{figure}[t]
\begin{center}
\includegraphics[width=1.0\columnwidth]{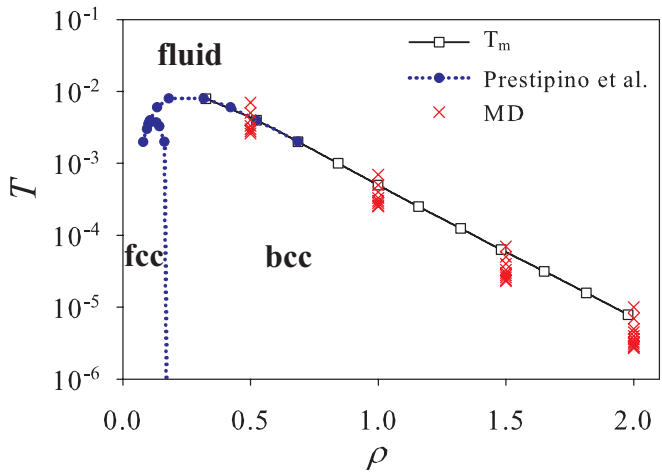}
\caption{
State points at which MD simulations were  performed (crosses). 
Squares with solid line and filled circles with dotted line
are the solid-fluid phase boundary obtained numerically 
by us~\cite{Ikeda2011,Ikeda_I} and Prestipino {\it et
 al.}~\cite{Prestipino2005}, respectively.  
The melting and freezing lines are indistinguishable at this scale. 
}
\vspace*{-0.3cm}
\label{phase}
\end{center}
\end{figure}

We investigate the dynamics of the one-component GCM using a molecular dynamics (MD) simulation in the $NVT$ ensemble 
with a Nos\'{e} thermostat.  
The system is a cubic cell and a periodic boundary condition is imposed. 
A time-reversible integrator, similar to the velocity-Verlet method, is
used with a potential cut-off at $r=5\sigma$~\cite{Frenkel2001}.  
Hereafter, $\sigma$, $\epsilon/\kb$, and $\sigma(m/\epsilon)^{1/2}$ are
taken as the units of the length, temperature, and time, respectively. 
The time step is fixed at 0.2, which is sufficiently small to conserve
the Nos\'{e} Hamiltonian during the long simulation runs.  
We focus on the four densities, $\rho = 0.5$, $1.0$, $1.5$, and $2.0$, 
and perform the MD simulations for various temperatures in the vicinity
of the melting temperature $T_m$. 
The state points which we performed simulation are shown in
Fig.~\ref{phase} along with the solid-fluid phase boundary
line~\cite{Prestipino2005,Ikeda2011,Ikeda_I}.  
As discussed in detail in the previous study~\cite{Ikeda_I}, 
the melting temperature, $T_m$, at the high density regime $\rho \gtrsim
1$ obeys an asymptotic scaling $\log T_m \propto -\rho^{2/3}$ which was
originally conjectured by Stillinger~\cite{Stillinger1976}.
For all densities which we study, thermodynamically stable crystalline structure is bcc~\cite{Ikeda2011,Ikeda_I}.
We run the simulations for the total run time always 50 times longer than
the structural relaxation time. 
For example,  the simulation time was $t_{sim} = 10^7$ for the lowest temperature 
at $\rho = 2.0$. 
This was confirmed to be sufficiently long to neglect aging effect.   
The first half of the simulation run was used for the equilibration and 
we used the trajectories of the second half for the analysis of the
stationary dynamics. 
For each state point, five independent runs are performed and the
results are obtained by averaging over those trajectories in order 
to improve the statistics.  
Configurations obtained from the high temperature simulation were used  
as the initial configurations.
The system size is fixed at $N=3456$.   
The simulations for $N=2000$ and $9826$ confirmed  that the finite-size effect is negligible.  

\subsection{Mode Coupling Theory}

In this work, we compare our simulation results for dynamics of the high
density GCM in the supercooled state with the prediction of MCT. 
In the context of the glass transition, MCT is commonly expressed 
as a set of the self-consistent nonlinear equations for 
correlation functions. 
These correlation functions are the intermediate
scattering function (the correlation of the collective density),
$F(k,t) \equiv \ave{\delta \rho(\vec{k},0) \delta \rho(-\vec{k},t)}/N$, 
where $\delta \rho(\vec{k},t)$ is the $k$-dependent density fluctuation,
and the self intermediate scattering function or the correlation
of the single particle density, 
$F_s(k,t)\equiv \ave{\rho_s(\vec{k},0) \rho_s(-\vec{k},t)}$,
where $\rho_s(\vec{k},t)$ is the density of a single particle. 
The time evolution of $F(k,t)$ is given by the generalized Langevin equation 
\begin{eqnarray}
 \begin{aligned}
\Omega^{-2}(k) \ddot{F}(k,t) + F(k,t) + \int^t_0\!\! ds \ M(k,t-s) \dot{F}(k,s) = 0,
 \end{aligned}
\label{eq:mctF}
\end{eqnarray}
where $\Omega(k)= \sqrt{\kb T k^2/mS(k)}$ is
the frequency term. 
$S(k)= F(k,t=0)$ is the static structure factor.
$M(k,t)$ is the memory kernel which, according to MCT, is approximated as   
\begin{eqnarray}
 \begin{aligned}
M(k,t) = \frac{\rho S(k)}{2k^2} \int\!\! \frac{d\vec{q}}{(2 \pi)^3}
 V_{\vec{k}}^2(\vec{q},\vec{k}-\vec{q}) F(q,t) F(|\vec{k}-\vec{q}|,t).
 \end{aligned}
 \label{mem}
\end{eqnarray}
Here $V_{\vec{k}}(\vec{q},\vec{p}) \equiv \{ \vec{k}\cdot\vec{q}c(q) + \vec{k}
\cdot\vec{p}c(p)\}/k$ is the vertex, where
$c(k)=\{1-1/S(k)\}/\rho$ is the direct correlation function. 
In Eq.~(\ref{mem}), we neglect the short time contribution for the
memory kernel, which does not affect the slow dynamics. 
MCT predicts that $F(k,t)$ undergoes
the ergodic-nonergodic transition at a finite temperature, $T_c$, 
below which 
$\lim_{t\rightarrow \infty}F(k,t)= F_{\infty}(k)$ remains finite. 
$F_{\infty}(k)$ is referred to as the nonergodic parameter. 
The nonergodic parameter and $T\mct$ can be evaluated by taking 
$t \to \infty$ of Eqs.~(\ref{eq:mctF}) and (\ref{mem}), which is 
expressed as
\begin{equation}
 \begin{aligned}
\frac{F_{\infty}(k)/S(k)}{1 - F_{\infty}(k)/S(k)} = M_{\infty}(k),   
 \end{aligned}
\label{eq:nep}
\end{equation}
where $M_{\infty}(k)$ is the long time limit of the memory kernel. 
As the temperature approaches to $T\mct$ from above, 
MCT first predicts that $F(k,t)$ exhibits
the two-step relaxation behavior characterized by a finite plateau and 
the slow structural relaxation. 
The height of the plateau is identical to $F_{\infty}(k)$ at $T=T\mct$.
The structural relaxation or the alpha relaxation time, $\tau_{\alpha}$, 
increases and eventually diverges at $T\mct$. 
MCT predicts that the increase of $\tau_{\alpha}$ is given by a power
law $\tau_{\alpha} \sim |T-T\mct|^{-\gamma}$, where $\gamma$ is a
system-dependent parameter which can be evaluated from the MCT equation. 

Likewise, the MCT equation for the self intermediate scattering function,
$F_s(k,t)$, is written in the same form as Eq.~(\ref{eq:mctF}), but with 
the frequency term $\Omega_s(k)= 
\sqrt{\kb T k^2/m}$ instead of $\Omega(k)$ and the self memory kernel 
\begin{equation}
\begin{aligned}
M_s(k,t) 
\!= \!\frac{\rho}{2k^2}\! \int\!\!\frac{d\vec{q}}{(2 \pi)^3}
\left\{\! \frac{\vec{k} \cdot \vec{q}}{k}c(q)\! \right\}^2 \!\!
F_s(q,t) F(|\vec{k}-\vec{q}|,t)
\end{aligned}
  \label{mems}
\end{equation}
instead of $M(k,t)$ in Eq.~(\ref{mem}). 
The MCT equation for $F_s(k,t)$ undergoes the nonergodic transition
exactly at the same temperature, $T\mct$, as for $F(k,t)$, 
at least for most model systems studied in the past (see Ref.~\cite{Voigtmann2010}
for exceptions). 
By taking the small $k$-limit of the MCT equation for $F_s(k,t)$, we can
also construct the self-consistent equation for the mean square
displacement $\ave{R^2(t)}$. 
MCT predicts that the self-diffusion coefficient $D \equiv
\lim_{t\rightarrow \infty}\ave{R^2(t)}/6t$ follows the power law
$D \sim |T-T\mct|^{\gamma}$ and vanishes at $T\mct$. 
Note that the power law exponent $\gamma$ is identical with that for $\tau_{\alpha}$.
In addition to the MCT nonergodic transition and power law of the transport
coefficients, MCT predicts many important dynamical properties 
such as the dynamic scaling known as von Schweidler's law at the
plateau regime (the beta regime) and the time-temperature superposition at
the alpha relaxation regime~\cite{Binder2005}. 

In order to solve the MCT equations, the static structure factor, $S(k)$, is required as an input. 
We used $S(k)$ obtained directly from simulations.
For the numerical integration of Eq.~(\ref{mem}) and (\ref{mems}), 
we employed equally spaced 400 grids with the grid spacing $\Delta k = 0.16$.

\section{Crystallization}
\begin{figure}[t]
\begin{center}
\includegraphics[width=1.0\columnwidth]{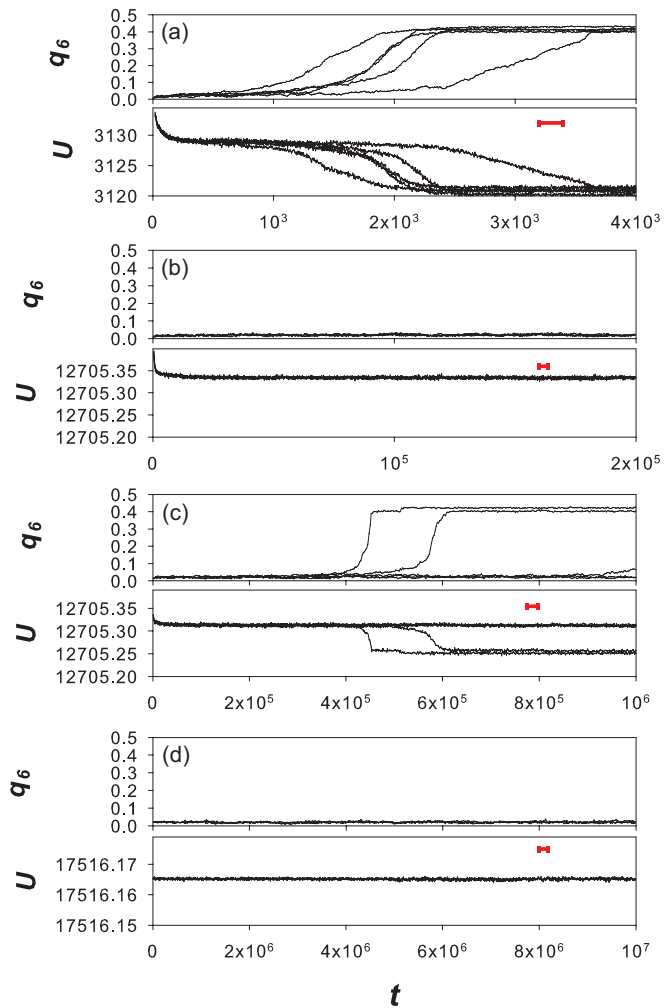}
\caption{
The time dependence of the bond order parameter $q_6$ and potential energy $U$ 
of the representative trajectories measured from the time when the
 system is prepared. 
(a) $\rho=0.5$, $T=2.5 \times 10^{-3}$, (b) $\rho=1.5$, $T=2.6 \times
 10^{-5}$,  (c) $\rho=1.5$, $T=2.3 \times 10^{-5}$, and (d) $\rho=2.0$, $T=2.93 \times 10^{-6}$. 
The short bold line in each figure indicates the time scale of $\tau_\alpha$. 
}
\vspace*{-0.3cm}
\label{pot}
\end{center}
\end{figure}

Ordinary simple atomic fluids nucleate to form crystals quickly as 
the temperature is lowered below the melting point.
In this section, we analyze the crystal nucleation dynamics of the high
density GCM and show that the nucleation rate systematically decreases 
as the density increases. 
In order to monitor the crystallization from the homogeneous fluid phase, 
we use the potential energy $U$ and the bond order parameter $q_6$~\cite{Steinhardt1983}. 
The bond order parameter is defined by 
\begin{equation}
 q_6 \equiv \frac{1}{N}\sum_{i=1}^{N} q_6(i),
\end{equation}
where $q_l(i)$ is the $l$-th bond order parameter of the $i$-the
particle defined by 
\begin{eqnarray}
q_{l}(i) = \sqrt{\frac{4 \pi}{2l + 1} \sum_{m=-l}^l |q_{lm}(i)|^2}. 
\label{eq:qli}
\end{eqnarray}
Here $q_{lm}(i)$ is the complex bond parameter of the $i$-th particle
given by 
\begin{eqnarray}
q_{lm}(i) = \frac{1}{N_b(i)} \sum_{j=1}^{N_b(i)} Y_{lm}(\vec{R}_i-\vec{R}_j), 
\label{eq:qlm}
\end{eqnarray}
where 
$\vec{R}_i$ is the position of the $i$-th particle,
$N_b(i)$ is the number of nearest neighbor particles around the
$i$-th particle, and
$Y_{lm}(\vec{r})$ is the spherical harmonic function of the degree $l$
and the order $m$. 

$q_6$ is zero in the fluid phase and $q_6\approx 0.5$ for a
perfect bcc crystal~\cite{Steinhardt1983}.  
In Fig.~\ref{pot}, we show $q_6$ and $U$ of the five representative trajectories 
as a function of the lapse of time measured from the moment when the system is prepared.
At a relatively low density $\rho=0.5$ and 
temperature just below the melting point $T=2.5\times 10^{-3}$ (Fig.~\ref{pot} (a)),  
one observes that $q_6$'s of all five trajectories abruptly increase from
zero to a finite value and concomitantly $U$'s decrease.
These behaviors are the hallmark of the crystal nucleation. 
This figure shows that the nucleation initiates only after the lapse of
time several times longer than the structural relaxation time
$\tau_{\alpha}$ 
 which is indicated by the short bold lines in the figures
(the precise definition and compiled data set of $\tau_{\alpha}$ are given in Sec.~\ref{sec:Glassy Dynamics}).  
The degree of supersaturation defined by $\Delta = 1 - T/T_m$ at this state point is 0.43. 
Next, we look at the higher density $\rho=1.5$.  
Five runs of $q_6$ and $U$ at $T=2.6 \times 10^{-5}$ are shown in
 Fig.~\ref{pot} (b). 
Despite of the deeper supersaturation ($\Delta = 0.55$) and much longer
 simulation runs (over 40 $\tau_{\alpha}$) than Fig.~\ref{pot} (a), 
$q_6$ and $U$ do not show any sign of nucleation.
Decreasing the temperature further to $T=2.3 \times 10^{-5}$ where
 $\Delta = 0.6$ (Fig.~\ref{pot} (c)), one eventually observes the crystallization for the two
 out of five trajectories.
Note that it takes decades of the structural relaxation time 
(which itself also increases with the degree of supersaturation) before
 the precipitous nucleation takes place.  
At even higher density $\rho=2.0$, all five trajectories fail to
 nucleate even at a very low temperature $T=2.93 \times 10^{-6}$ with
 the similar degree of the supersaturation, $\Delta = 0.6$, 
for the whole simulation runs.

\begin{figure}[t]
\begin{center}
\includegraphics[width=1\columnwidth]{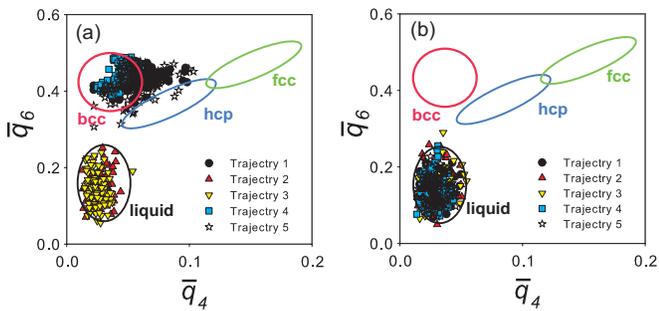}
\caption{
The $\bar{q}_4$-$\bar{q}_6$ correlation map for the configurations
 obtained at the end of all the five simulation runs 
at
$(\rho, T) = (1.5,  2.3 \times 10^{-5})$ (left panel) 
and $(\rho, T) = (2.0,  2.93 \times 10^{-6})$ (right panel). 
Four circles are the characteristic distribution for the bcc,
 hcp, fcc crystal, and fluid phase.  
}
\vspace*{-0.3cm}
\label{qmap}
\end{center}
\end{figure}
In order to ensure that the nucleated samples
are unambiguously the bcc crystal and that samples which failed to nucleate
remain in the homogeneous fluid phase, 
we evaluate new parameters
which were recently introduced by Lechner {\it et al.}~\cite{Lechner2008}. 
They have used the two averaged bond order parameters $\bar{q}_4(i)$ and $\bar{q}_6(i)$ 
and demonstrated that the correlation map of them
improves ability to determine the crystalline structures~\cite{Lechner2008,Kawasaki2010}.  
The averaged bond order parameter is defined by replacing $q_{lm}(i)$ in
Eq.~(\ref{eq:qli}) with the averaged value $\bar{q}_{lm}(i)$ defined by
\begin{eqnarray}
\bar{q}_{lm}(i) = \frac{1}{\tilde{N}_b(i)} \sum_{k=0}^{\tilde{N}_b(i)} q_{lm}(k),  
\end{eqnarray}
where $q_{lm}(k)$ is given by Eq.~(\ref{eq:qlm}) and the sum runs from
$k$ over all $\tilde{N}_b(i)$ neighbors of the $i$-th particles,
including the $i$-th particle itself ($k=0$ in the sum). 
In Fig.~\ref{qmap}, we placed all $\bar{q}_4(i)$ and $\bar{q}_6(i)$
($i=1,2,\cdots, N$) in the correlation map for 
the configurations obtained at the end of simulation
runs of the two state points $(\rho, T) = (1.5, 2.3 \times 10^{-5})$ 
and $(\rho, T) = (2.0, 2.93 \times 10^{-6})$. 
The four circles represent the characteristic areas for
the bcc, hcp, fcc crystals, and fluid phase~\cite{Lechner2008}.   
The results for $(\rho, T) = (1.5, 2.3 \times 10^{-5})$ show 
that the two trajectories remain in the fluid phase whereas the rest
formed  the bcc crystal.
It is clear that no other structures are formed in the course of the
simulations. 
Note that the results for the three trajectories which nucleated 
slightly deviate from the bcc region, which we presume 
is due to defects or imperfectness of the obtained crystalline structures. 
On the other hand, all the five trajectories for 
$(\rho, T) = (2.0, 2.93 \times 10^{-6})$ do not show any hint of
crystal nucleation and the configurations remain completely disordered. 
Hereafter, we focus on the densities $\rho=1.5$ and 2.0 
because the crystal nucleation is sufficiently slow that canonical
glassy dynamics are observed.

\section{Glassy Dynamics}\label{sec:Glassy Dynamics}

\subsection{Structural functions}

\begin{figure}[t]
\begin{center}
\includegraphics[width=1.0\columnwidth]{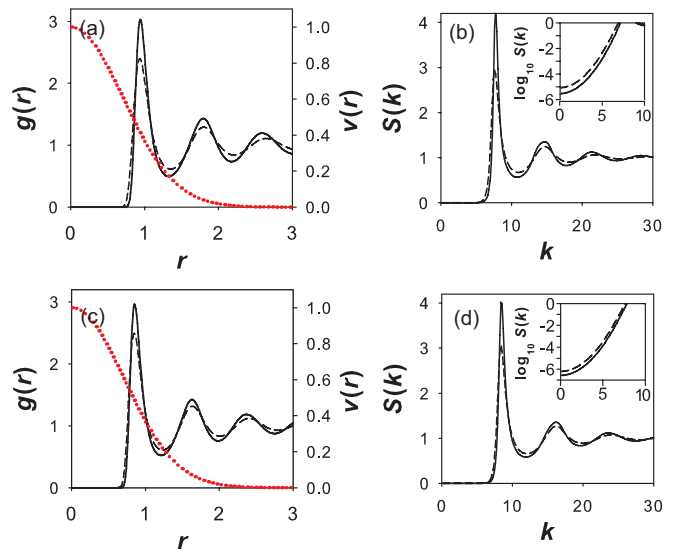}
\caption{
The radial distribution function $g(r)$ (left panels)
and the static structure factors $S(k)$ (right panels). 
(a) and (b) are for $\rho = 1.5$ at $T = 7.0 \times
 10^{-5}$ (dashed line) and $T = 2.4 \times 10^{-5}$ (solid line). 
(c) and (d) are for $\rho = 2.0$ 
at $T = 7.0 \times 10^{-6}$ (dashed line) and $T = 2.93 \times 10^{-6}$ (solid line). 
The insets of (b) and (d) are the closeup of $S(k)$ at small $k$'s in the semilog plot.
Dotted lines in (a) and (c) are the bare potential $v(r)$.
}
\vspace*{-0.3cm}
\label{sk}
\end{center}
\end{figure}

Before discussing the slow dynamics in the supercooled state, 
we summarize the fluid structures of the high density GCM 
to demonstrate the difference from those of conventional model glassformers. 
In Fig.~\ref{sk}, we plot 
the radial distribution functions $g(r)$ and static structure factors $S(k)$
of the GCM for $\rho=1.5$ and 2.0 near and below the melting temperatures.
Both $g(r)$ and $S(k)$ show typical behaviors of dense fluids 
characterized by the prominent peaks near 
the position and the wavevector corresponding to the first coordination shell. 
Their peak heights increase as the temperature decreases.
As density increases from $\rho=1.5$ to 2.0,
the peak position of $g(r)$ shifts from $r=0.94$ to 0.85 
and for $S(k)$ from $k=7.8$ to 8.4.
The noticeable feature of the high density GCM 
is that the tail of the potential $v(r)$ stretches beyond the first
coordination shell, as demonstrated in Figs.~\ref{sk} (a) and (c). 
This considerable overlap of particles imparts the character of the
long-ranged interaction systems to the high density GCM. 
The long-ranged nature also appears 
as the anomalously small $S(k)$ at small wavevectors. 
The insets of Figs.~\ref{sk} (b) and (d) show that $S(k\approx 0)$, or the
compressibility, is far smaller than the other model fluids at
compatible supersaturations,  
implying  that the density fluctuations at
large length scales are strongly suppressed. 
This is the common features of the long-range interacting
systems. 
A well-known example is the one component classical
plasma~\cite{Ichimaru1982}, where $S(k)$ vanishes at $k \to 0$.  
More detailed analysis of the simulation results for the structural 
functions and comparisons with the predictions of the liquid state theory have been
reported in Ref.~\cite{Ikeda_I}.  
$S(k)$'s obtained here are used in the MCT analysis discussed below.

\subsection{Mean square displacement and self intermediate scattering function}

\begin{figure*}[t]
\begin{center}
\includegraphics[width=1.7\columnwidth]{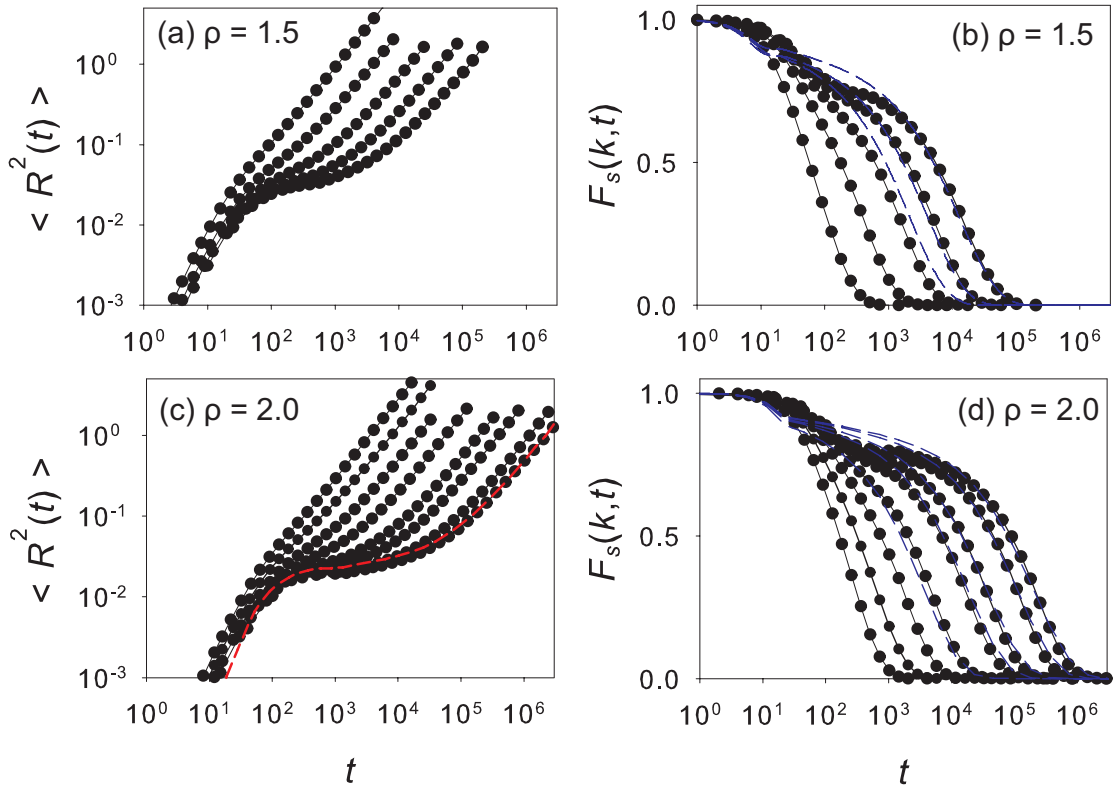}
\caption{
$\ave{R^2(t)}$ ((a) and (c))  and 
$F_s(k_{\max},t)$ ((b) and (d)). 
The filled circles are simulation results for
$\rho = 1.5$ and from left to right, $T\times 10^5=7$, $4$, $3$, $2.6$
 and $2.4$ (upper panel), 
and for $\rho = 2.0$ and from left to right, $T\times 10^6=10$, $7$,
 $5$, $4$, $3.4$, $3.2$, $3$ and $2.93$ (lower panel). 
The dashed line in (c) is the mean square displacement of the KA model
at $T=0.475$~\cite{Kob1994} shifted to fit with the GCM's result
at the lowest temperature at long times (see text). 
The dashed lines in (b) and (d) are the MCT solutions obtained using the 
same reduced temperatures, $\varepsilon$, as those for the simulation data.
}
\vspace*{-0.3cm}
\label{msdfskt}
\end{center}
\end{figure*}
\begin{figure}[t]
\begin{center}
\includegraphics[width=1.0\columnwidth]{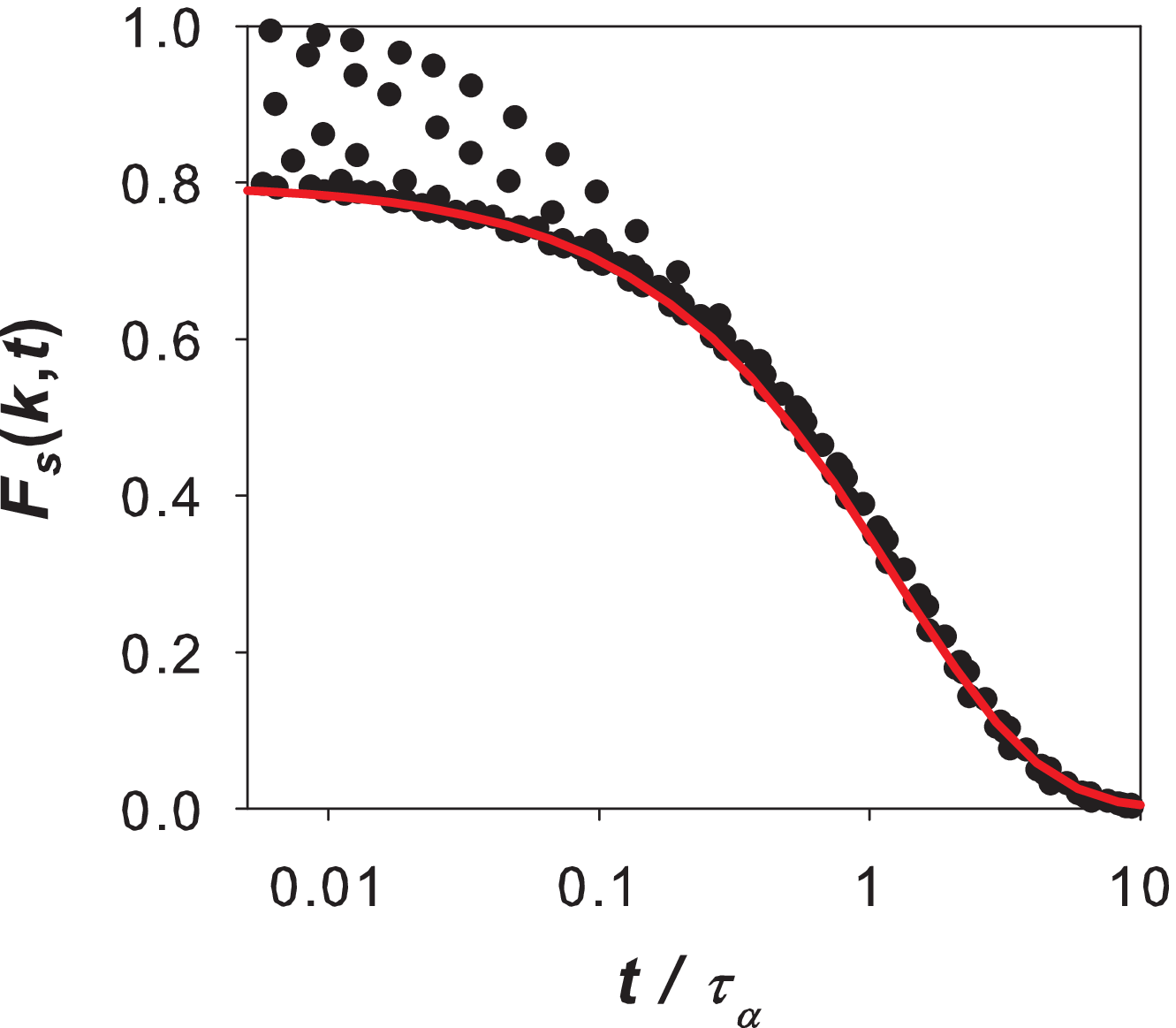}
\caption{
Same as Fig.~\ref{msdfskt} (d) but plotted against $t$ scaled by $\tau_{\alpha}$.
Filled circles are the simulation results for $\rho = 2.0$ and $T\times
 10^6=5$, $4$, $3.4$, $3.2$, $3$, $2.93$ from right to left.  
The solid line is a fit by a stretched exponential function. 
}
\vspace*{-0.3cm}
\label{ttsp}
\end{center}
\end{figure}

In this subsection, we evaluate various dynamic quantities 
and observe their slow dynamics, focusing on the trajectories 
which did not crystallize even when deeply supercooled.
The mean square displacement 
$\ave{R^2(t)} \equiv N^{-1}\sum_{i=1}^{N} \ave{|\vec{R}_i(t) - \vec{R}_i(0)|^2}$, 
the self intermediate correlation function $F_s(k,t)$, 
and the intermediate correlation function $F(k,t)$
are evaluated for the densities $\rho=1.5$ and 2.0. 
Fig.~\ref{msdfskt} shows 
$\ave{R^2(t)}$ and $F_s(k,t)$ at several temperatures well
below the melting temperature. 
These figures clearly display the canonical behaviors of the supercooled
liquids near the glass transition point.  
For $\rho=1.5$, we could not observe the
glassy dynamics below $T=2.4\times 10^{-5}$ because the crystallization intervened. 
At $\rho=2.0$, all trajectories did not crystallize down to the lowest temperature which we accessed. 
In Figs.~\ref{msdfskt} (a) and (c), 
one observes that, as the temperature is lowered, $\ave{R^2(t)}$ develops the long plateau
regimes followed by the usual diffusive behaviors $\ave{R^2(t)}\propto t$ at longer times. 
The appearance of the plateau signals the formation of a cage of a 
particle surrounded by its neighbors and is the hallmark of the supercooled
fluid near the glass transition point. 
The value of $\sqrt{\ave{R^2(t)}}$ at the plateau region is 
a measure of the sizes of the cages. 
They are about $\sqrt{\ave{R^2(t)}}\approx $0.17 for $\rho=1.5$ and 0.14
for $\rho=2.0$.
These values are slightly smaller than the values for conventional model
glassformers. 
For example, $\sqrt{\ave{R^2(t)}}\approx 0.2$ for the Kob-Anderson Lennard-Jones
mixture (KA model)~\cite{Kob1994}.

In Fig.~\ref{msdfskt} (c) and (d), we plot $F_s(k=k_{\max},t)$ for several temperatures,  
where  $k_{\max}$ is the wavevector where $S(k)$ show the maximum peak.  
$F_s(k_{\max},t)$ relaxes exponentially at high temperatures.  
As the temperature decreases, a plateau with a finite height appears
and it stretches over longer times as the temperature decreases
further, while the plateau height remains almost constant. 
This two-step relaxation behavior is another hallmark of the slow
dynamic near the glass transition point. 
The terminal relaxation following the plateau is called the 
structural or alpha relaxation. 
We define the structural relaxation time $\tau_{\alpha}$ by $F_s(k_{\max},t=\tau_{\alpha}) = e^{-1}$. 
In Fig.~\ref{ttsp}, we plot $F_s(k_{\max},t)$ against the time scaled by
$\tau_{\alpha}$.  
The result shows that relaxation curves are collapsed at the alpha relaxation regime. 
This is the universal property of the glassy systems known as the
time-temperature superposition (TTS)~\cite{Binder2005}.  
Furthermore, the all curves where TTS holds are fitted by a stretched exponential function 
$e^{-(t/\tau_{\alpha})^{\beta}}$ with the exponent $\beta \approx  0.8$. 
This value is comparable with that for the KA model~\cite{Kob1994} and for the hard sphere mixture~\cite{Foffi2004}.

\begin{figure}[t]
\begin{center}
\includegraphics[width=1\columnwidth]{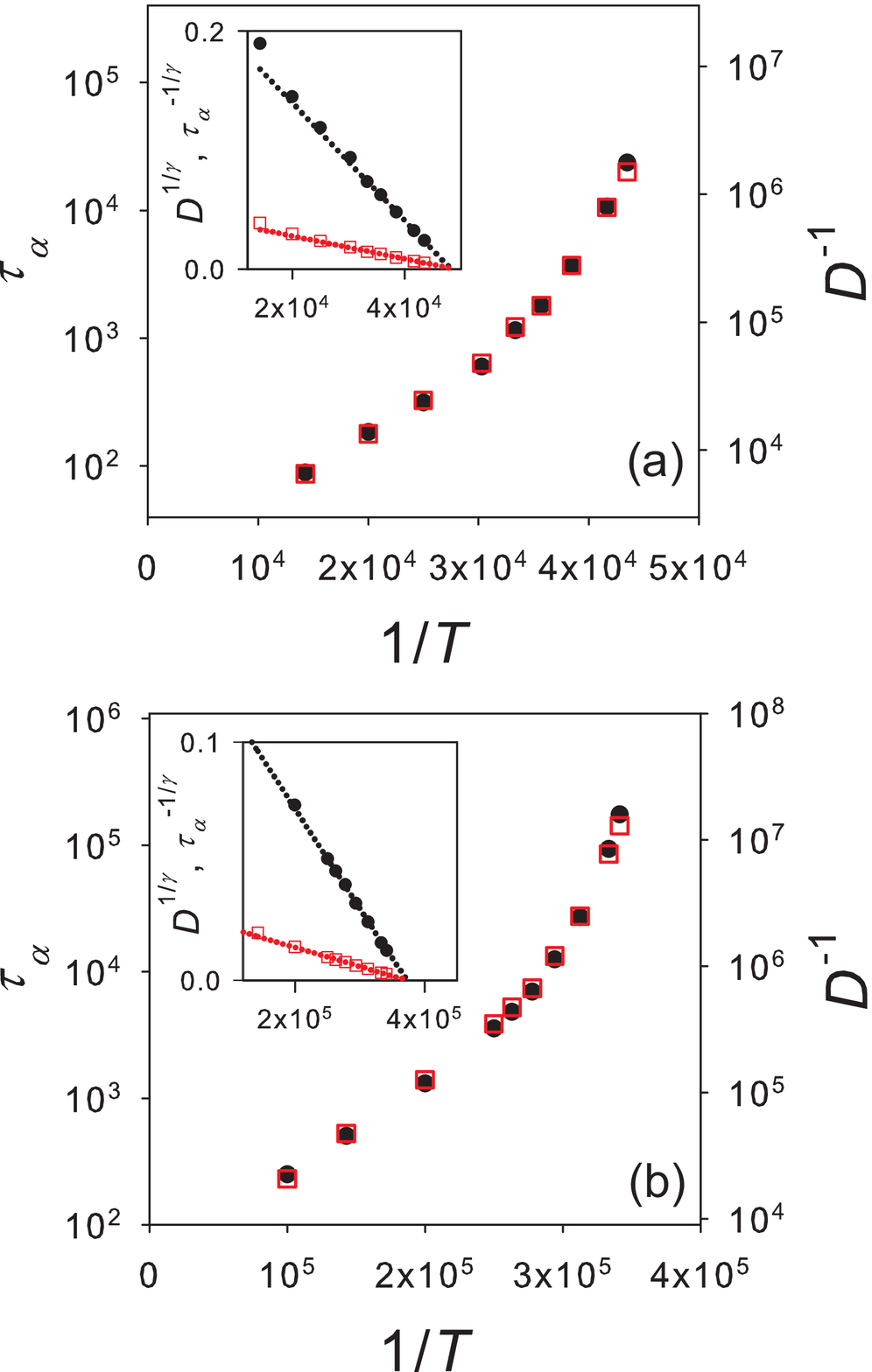}
\caption{
The temperature dependence of the structural relaxation time (filled circles) 
and the inverse of the diffusion coefficient (empty squares) for (a) $\rho = 1.5$ and (b) $\rho = 2.0$. 
Inset: $\tau_{\alpha}^{-1/\gamma}$ and $D^{1/\gamma}$ as a function of
 inverse temperature, where $\gamma$ is fixed to 2.7. 
}
\vspace*{-0.1cm}
\label{angell}
\end{center}
\end{figure}
In Fig.~\ref{angell}, the structural relaxation time $\tau_{\alpha}$ and
the self diffusion constant defined by $D \equiv \lim_{t \to \infty}
\ave{R^2(t)}/6t$ are plotted against the inverse temperature.
We plotted $D^{-1}$ and adjusted its ordinate so that the data collapses
with $\tau_{\alpha}$
at high temperatures.
For both densities, $\rho=1.5$ and $2.0$,
$\tau_{\alpha}$ and $D^{-1}$ drastically increase 
as the temperature is lowered. 
Both data almost collapse to each other for the whole temperatures 
except for the slight deviation at the lowest temperature.
As we shall discuss later, this is the direct reflection of a weak
violation of the Stokes-Einstein relation.

So far, all simulation data show no sign of peculiarity in 
the slow dynamics of the high density GCM at the qualitative level. 
They are all similar to conventional model glassformers. 
In order to assess the properties of the high density GCM more
quantitatively, we compare the simulation results with the predictions of MCT. 
For this purpose, we solve the MCT equations
Eqs.~(\ref{eq:mctF})--(\ref{mems}) by numerically integrating the
equations in a self-consistent manner. 
As inputs, we used $S(k)$ obtained numerically in the previous subsection.
First, we compute the MCT transition temperature $T\mct$ by solving Eq.~(\ref{eq:nep}). 
The results are $T\themct = 2.66 \times 10^{-5}$  and $3.17 \times 10^{-6}$ for $\rho=1.5$ and 2.0, respectively. 
Here, we denote the transition temperature as $T\themct$ in order to emphasize that they
are obtained by solving the MCT equations.
The exponent $\gamma \approx 2.7$ is also obtained from the MCT solutions.

\begin{figure}[t]
\begin{center}
\includegraphics[width=1\columnwidth]{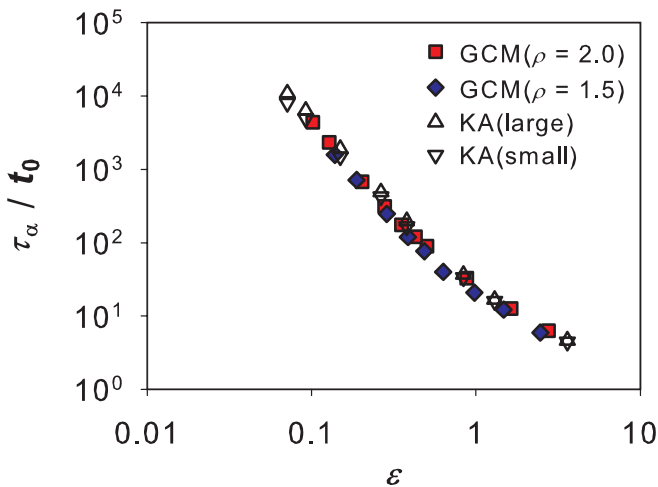}
\caption{
$\tau_{\alpha}/t_0$ as a function of the reduced temperature
 $\varepsilon$ for the GCM and KA model. 
$t_0$ is the short-time relaxation time defined by $F_s(k_{\max},t_0)=0.95$. 
}
\vspace*{-0.1cm}
\label{scale}
\end{center}
\end{figure}

MCT predicts that both the self-diffusion coefficient and the structural 
relaxation time follow the power law $D^{-1}, \tau_{\alpha} \propto
|T-T\mct|^{-\gamma}$ with the same parameters $\gamma$ and $T\mct$.  
We fitted $D^{-1}$ and $\tau_\alpha$ obtained by simulation with this 
MCT power law, using $T\mct$ as a fitting parameter. 
We denote it as $T\simmct$. 
By plotting $D^{\gamma}$ and $\tau_{\alpha}^{-\gamma}$ against $T^{-1}$,
we found that they both vanish at the same temperature and we identified 
$T\simmct = 2.07 \times 10^{-5}$ and $2.68 \times 10^{-6}$ for $\rho =
1.5$ and 2.0, respectively (see the insets of Fig.~\ref{angell}). 
In Fig.~\ref{scale}, we replotted $\tau_{\alpha}$ in Fig.~\ref{angell} 
using $\varepsilon \equiv 1 - T/T\simmct$ instead of $1/T$. 
The results for the KA model~\cite{Kob1994} are also
plotted. 
These data are scaled by a time unit, $t_0$, defined by a relaxation time
at the short time scale, $F_s(k_{\max},t= t_0) = 0.95$.
This figure shows that the relaxation times for both the GCM and KA model
ride on the MCT power law for the range of temperatures which the
simulation can access. 
Collapse of the data of two systems on a single power law is a reflection that 
the values of $\gamma$'s of both systems are close ($\gamma\approx 2.5$ for
the KA model~\cite{Flenner2005d}).
This figure also demonstrates that $\varepsilon$ is 
a good parameter to measure the distance from the onset of the
glassy slow dynamics for different systems. 
Hereafter, we refer to $\varepsilon$ as the reduced temperature.
In Fig.~\ref{msdfskt} (c), we plotted the simulation data of $\ave{R^2(t)}$ for the KA model at
$T=0.475$ by shifting the time unit in such a way that 
the long time diffusive regime collapses with the data for the GCM 
at $T=2.93 \times 10^{-6}$ and $\rho=2.0$ whose reduced temperature 
is about the same. 
Almost perfect collapse of the results for  two distinct systems
for the whole time window, including the short time ballistic behavior 
and the entry to the plateau regime, suggests that the slow
diffusive behavior of the high density GCM is qualitatively similar to
that of canonical glassformers at least above $T\simmct$, where
our MD simulation can access.

However, there are two noticeable differences between the high
density GCM and conventional model glassformers.
First, the MCT transition temperature obtained from fitting the
simulation data, $T\simmct$, is unprecedentedly close to the theoretical
prediction $T\themct$ for the GCM.
The agreement improves as the density increases.
The deviation of $T\simmct$ from $T\themct$ are only 32 \% for $\rho=1.5$ and 
20 \% for $\rho=2.0$.
It is in stark contrast with the KA model for which 
$T\simmct = 0.435$ and $T\themct = 0.92$ with the deviation of more
than 100\%~\cite{Kob2002,Flenner2005e}. 
The KA model at $T\themct$ is still a high-temperature fluid and $F_s(k,t)$ decays
exponentially without a sign of two-step relaxation. 
Contrarily, the GCM at $T\themct$ already lies deep in the region where
the plateau of $F_s(k,t)$ is well developed (see Fig.~\ref{msdfskt} (d)).
Considerable deviation of $T\simmct$ from $T\themct$ for conventional model
glassformers is known as one of serious drawbacks of MCT.
These deviations have been attributed to the effect of 
the activated processes in the ragged energy landscapes, which smears out
the clear-cut dynamical
transition~\cite{sastry1998,Brumer2004b,Mayer2006b,Bhattacharyya2008}. 
Second, the MCT parameters $T\mct$ and $\gamma$ obtained from fitting simulation data for
$\tau_{\alpha}$ match very well with that obtained from the data of $D^{-1}$. 
This is also in contrast with the model glassformers 
such as the KA model~\cite{Kob1994,Flenner2005e} and 
poly(bi)disperse hard spheres~\cite{Foffi2004,Kumar2006}, 
for which $T\simmct$ (or the transition density $\rho\simmct$) and $\gamma$ obtained from fitting the
simulation data vary depending on the observables ($\tau_{\alpha}$ or $D^{-1}$) 
and also on the components (large or small particles components of the
binary systems).
These variances are partly attributed to the presence of strong dynamic
heterogeneities which decouple the diffusion from the structural relaxation
time, as we shall discuss in the next subsection.

The direct evidence that MCT works better for the GCM than any other
model glassformers is the remarkable agreement of the simulated
$F_s(k,t)$ with the MCT prediction.
In Fig.~\ref{msdfskt} (b) and (d), we plotted the solutions of MCT
for exactly the same reduced temperatures $\varepsilon$ as the
simulation data. 
Only free parameter is the time unit, which is determined solely from the
short time dynamics. 
Long time behaviors of the MCT solution agree very well with the simulation
results. 
MCT also correctly predicts the exponent of the stretched exponential
relaxation $\beta$. 
The agreement is striking given that for other model glassformers, $\varepsilon$ 
(and sometimes the wavevectors as well) needs to be adjusted at each temperature 
to obtain a reasonable fit~\cite{Kob2002,Voigtmann2004} 
(an exception is the four-dimensional system~\cite{Charbonneau2010}).  

\begin{figure}[t]
\begin{center}
\includegraphics[width=1\columnwidth]{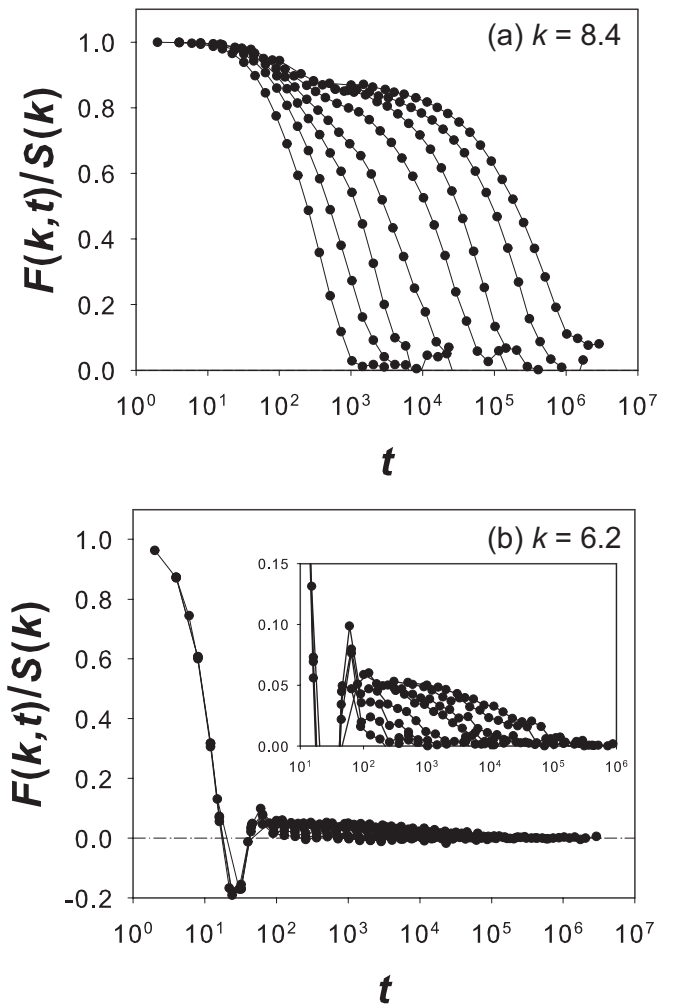}
\caption{
The intermediate scattering function at (a) $k=8.4$ and (b)
 $k=6.2$. 
For both panels, $\rho=2.0$ and the temperatures are, from left to right, 
$T \times 10^6 =$ 10, 7, 5, 4, 3.4, 3.2, 3 and 2.93.  
The inset in (b) shows a closeup of the weak and long tails of the main panel.
}
\vspace*{-0.1cm}
\label{fkt}
\end{center}
\end{figure}

\subsection{Intermediate scattering function}

Next, we look at the intermediate scattering function
$F(k,t)$. 
For conventional model glassformers, it is known that behavior of $F(k,t)$
is qualitatively the same as that of $F_s(k,t)$, except for the wiggly 
$k$-dependence of the nonergodic parameter for the former, 
reflecting the wiggly profiles of the static structure factor (see the discussion below). 
Contrarily,  for the high density GCM, $F(k,t)$ and $F_s(k,t)$ differ
from each other considerably. 
Fig.~\ref{fkt} shows $F(k,t)$ at two wavevectors. 
Fig.~\ref{fkt} (a) is the result at $k = k_{\max}(\approx 8.4)$ which is
the peak position of $S(k)$. 
There, the relaxation behavior of $F(k,t)$ is very similar to that
of $F_s(k,t)$, suggesting the relaxations of both functions at 
the interparticle length scales are dictated by the same relaxation mechanism.
Fig.~\ref{fkt} (b)  is the result at 
$k=6.4$, which corresponds to a slightly longer length scale 
than the interparticle distance.  
The relaxation of $F(k,t)$ is very fast and shows no sign of two
step relaxation. 
$F(k,t)$ almost fully relaxed at $t\sim 10$, which is much shorter than the
onset time of the caging where the plateau of $\ave{R^2(t)}$ appears 
(see Fig.~\ref{msdfskt}). 
The quick decays are followed by the phonon-like oscillations and very weak tails persisting
over the time scale of the structural relaxation time.
This tail vanishes at smaller $k$'s.
This behavior is in sharp contrast with the KA model, where the
relaxation time at small wavevectors is comparable with that at the
interparticle distance and the plateau heights remains finite down to
very small wavevectors~\cite{Gleim1998}.  
These results indicate that,  in the high density GCM,  
the large scale density fluctuations are decoupled from the slow
structural relaxation processes at the shorter length scales.  

\begin{figure}[t]
\begin{center}
\includegraphics[width=1\columnwidth]{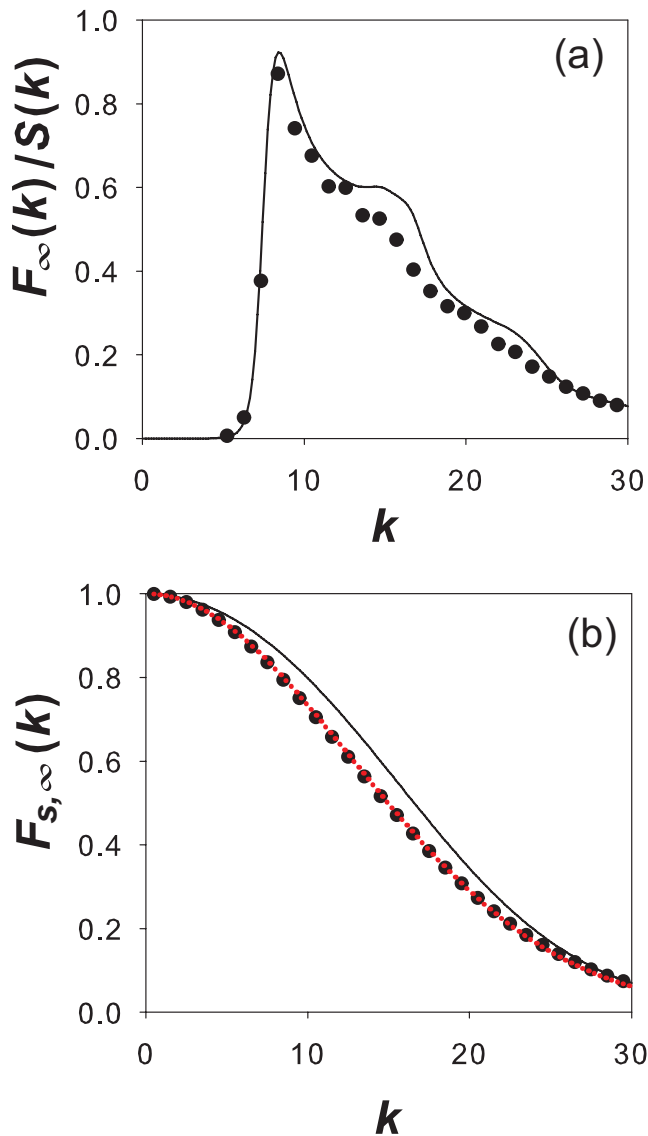}
\caption{
Nonergodic parameters for the collective part $F_{\infty}(k)/S(k)$
 (upper panel) and self part $F_{s,\infty}(k)$ (lower panel) of the 
intermediate scattering functions. 
Filled circles are the simulation data 
and solid lines are the MCT solutions. 
The dotted line in the lower panel is a fit by a Gaussian function.  
}
\vspace*{-0.1cm}
\label{nep}
\end{center}
\end{figure}

In order to see this qualitative difference of 
$F(k,t)$ of the GCM more clearly, 
we plot the $k$-dependence of the plateau heights, or the
nonergodic parameter, $F_{\infty}(k)$ and $F_{s,\infty}(k)$
together with the MCT predictions obtained from Eq.~(\ref{eq:nep}). 
In Fig.~\ref{nep}, we show $F_{\infty}(k)/S(k)$ and $F_{s,\infty}(k)$ at $\rho=2.0$
(filled circles) and the MCT predictions at the same density (solid lines). 
It beyond doubt demonstrates that MCT beautifully captures the vanishing
plateau and the decoupling between the self and collective dynamics at small
wavevectors. 
One observes that $F_{\infty}(k)/S(k)$ above $k_{\max}$
remains compatible with that of $F_{s,\infty}(k)$, while keeping a wiggly 
behavior characteristic of the
collective density fluctuations. 
Absence of slow dynamics 
at small $k$'s is a consequence of the anomalous structural properties inherent
in the high density GCM. 
In the previous subsection, we discussed that the static structure
factor at the small wavevectors, or the compressibility, 
is extremely small compared with those of ordinary model glassformers.
This makes the amplitude of the memory kernel at small $k$'s
negligibly small (see Eq.~(\ref{mem})).
Consequently the large scale fluctuations decouple from the 
fluctuations at the length scales of the interparticle distance which
trigger the glassy  slow dynamics.  
We argue that this decoupling between small and long length
scales should be commonly observed for the systems with small
compressibilities which are an universal feature of the
dense and long ranged interaction systems including the Coulomb
interaction systems as predicted in the framework of MCT~\cite{Shiroiwa2010}.

The nonergodic parameters in Fig.~\ref{nep} exhibit another subtle but
noticeable feature which may have relevance to fundamental problems of MCT as the mean field
description of the glass transition. 
Although MCT reproduces the overall behaviors of the nonergodic parameters
for both $F_{\infty}(k)/S(k)$ and $F_{s,\infty}(k)$, 
its prediction systematically overestimates the simulation results at
the intermediate wavevectors (in the range of, say, $5 \lesssim k \lesssim 20$). 
As shown in Fig.~\ref{nep} (b), we find that 
the simulation data for $F_{s,\infty}(k)$ is well fitted by a Gaussian
function, whereas the MCT nonergodic parameter has a small but
non-negligible shoulder which the Gaussian function can not fit. 
This shoulder is reminiscent of those observed in the MCT solution for
hard sphere glasses in large spatial dimensions~\cite{Schmid2010b, Ikeda2010}. 
There, we have found that the deviation from the Gaussian function for
$F_{s,\infty}(k)$ increases as the dimension $d$ increases. 
This observation has lead us to conclude that MCT is not rigorously
a bona fide mean field theory~\cite{Ikeda2010}.  
This glitch of MCT which we found in one of the mean field limits,
{\it i.e.}, the high $d$ limit, could also show up in another mean field limit, that is, 
the long-ranged interaction limit, which is realized in the high density
limit of the ultrasoft potential systems such as the GCM. 
This may explain the shoulder of the $F_{s,\infty}(k)$ in
Fig.~\ref{nep} (b). 
Remember that the anomalously small $S(k)$ at small $k$'s  is
also due to the long-ranged interaction. 
Interestingly, this small $S(k)$ 
may explain the anomalous shoulder of the MCT solution.
By artificially enhancing the amplitude of $S(k)$ at small $k$'s 
by a minute amount and plugging the modified $S(k)$ into the MCT equation, we find that 
the nonergodic parameter $F_{\infty}(k)$ at small $k$'s jumps from zero to 
finite values. 
At the same time, the shoulder of $F_{s,\infty}(k)$ at the intermediate wavevectors
disappears and MCT's $F_{s,\infty}(k)$ gets closer to the simulation results. 
This observation implies that the long range interaction affects the
static properties of the large length scales, which eventually amplifies
the putative non-Gaussian behaviors of the MCT solution.
A subtle interplay between the long and short length fluctuations 
may be quite common for the glass or/and jamming transition: 
For example, 
the the hyper-uniformity (vanishing $S(k)$ at small $k$)
and diverging radial distribution function at the contact length $r =\sigma$
are known to be 
the two facets of a universal character of the jamming transition~\cite{Torquato2003}.

\subsection{Violation of Stokes-Einstein relation}
\begin{figure}[t]
\begin{center}
\includegraphics[width=1\columnwidth]{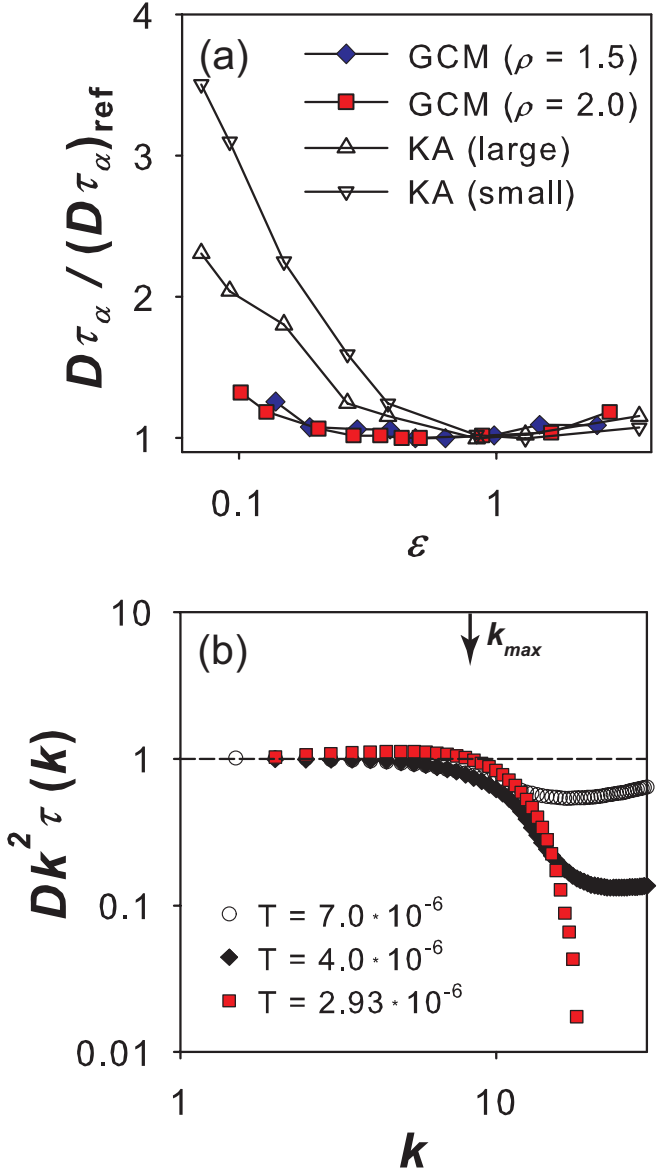}
\caption{
(a) Reduced-temperature dependence of $D\tau_{\alpha}$ at $\rho=1.5$ (diamonds) and 2.0 (squares). 
The results for the KA model are also plotted (triangles)~\cite{Kob1994}. 
All results are normalized by those at a high temperature $(D\tau_{\alpha})_{{\mbox{\scriptsize ref}}}$. 
(b) $Dk^2\tau(k)$ for $T\times 10^6=7.0$ (circles), 4.0 (diamonds)
 and 2.93 (squares) at $\rho=2.0$. The arrow indicates $k_{\max}$, the first peak of $S(k)$.
}
\vspace*{-0.1cm}
\label{tauk}
\end{center}
\end{figure}

For many glassformers, the Stokes-Einstein (SE) relation $D \approx T/\eta$, where 
$\eta$ is the shear viscosity, is violated near the glass transition
point and the violation is believed to be the manifestation of
spatially heterogeneous dynamics which grows as the temperature is
lowered~\cite{ediger2000}. 
Indeed, MCT can not capture the SE violation due to its mean field
character.
In this section, we show that the SE violation for the high density GCM is
suppressed. 
In Fig.~\ref{tauk} (a), we plot $D\tau_{\alpha}$ for $\rho=1.5$ and 2.0
normalized by the values at a high temperature, as a function of $\varepsilon$.
Note that $D\tau_{\alpha}$ instead of $D\eta$ has been plotted, because
$\eta$ and $\tau_{\alpha}$ are roughly proportional to each other.
In the same figure, we have also plotted the results for the large and
small particles for the KA model~\cite{Kob1994}. 
It is obvious that the variations of $D\tau_{\alpha}$
for the GCM is much weaker than that of the KA model.
Similar suppression of the SE violation was observed in the four-dimensional
hard sphere system~\cite{Charbonneau2010}.  

$\tau_{\alpha}$ was defined by $F_s(k, \tau_{\alpha})= e^{-1}$ at $k = k_{\max}$.
In order to study the length scales which are relevant to the SE violation,
we generalize the structural relaxation time to the $k$-dependent form, 
$\tau(k)$, defined by $F_s(k,\tau(k)) = e^{-1}$. 
Note that $\tau_{\alpha} = \tau(k_{\max})$.
In the small wavevector limit, 
the self intermediate scattering function behaves as 
$F_s(k,t) =e^{-Dk^2t}$. Therefore, $\tau(k) \sim 1/Dk^2$ as
$k\rightarrow 0$. 
In the opposite limit, the system 
should behave as an ideal gas, so that
$F_s(k,t) = {\displaystyle \e^{-\kb T k^2t^2/m}}$. 
Thus, $\tau(k) \propto 1/k$ as $k \rightarrow \infty$~\cite{hansen1986}. 
Fig.~\ref{tauk} (b) shows $Dk^2 \tau(k)$ as a function of $k$ for $\rho=2.0$
and several temperatures.  
Similar analysis for the KA model has been done by Flenner {\it et al}.~\cite{Flenner2005e}. 
At a high temperature $T=7.0\times10^{-6}$ where 
the two-step relaxation of $F_s(k,t)$ is set off (see
Fig.~\ref{msdfskt} (d)), $Dk^2\tau(k)$ is nearly constant and almost 1 at small wavevectors 
up to $k_{\max}$.
It then decreases as $k$ increases further, followed by a turn over to a
mildly increasing function. 
The decrease is a reflection of the vanishing of the cages at
length scales  shorter than the interparticle distance. 
The increase at larger $k$ is a crossover to the ideal gas limit where
$Dk^2 \tau(k) \propto k$. 
The qualitative behavior remains unchanged at $T = 4.0\times10^{-6}$, 
but the drop at $k \gtrsim k_{\max}$ is more pronounced, 
reflecting the stronger cage effect at lower temperatures.
At the lowest temperature $T=2.93\times10^{-6}$ which corresponds to
about $\varepsilon \approx 0.075$, the drop at $k \gtrsim 
k_{\max}$ is more dramatic. 
Furthermore, slight positive bump at $3
\lesssim k \lesssim k_{\max}$ is observed. 
This deviation corresponds to a weak SE violation observed in Fig.~\ref{tauk} (a). 
This behavior is noticeably different from that for the KA model
for which $Dk^2 \tau(k)$ significantly increases as $k$ increase before
dropping near $k_{\max}$~\cite{Flenner2005e}. 

\subsection{Non Gaussian dynamics}

\begin{figure}[t]
\begin{center}
\includegraphics[width=1\columnwidth]{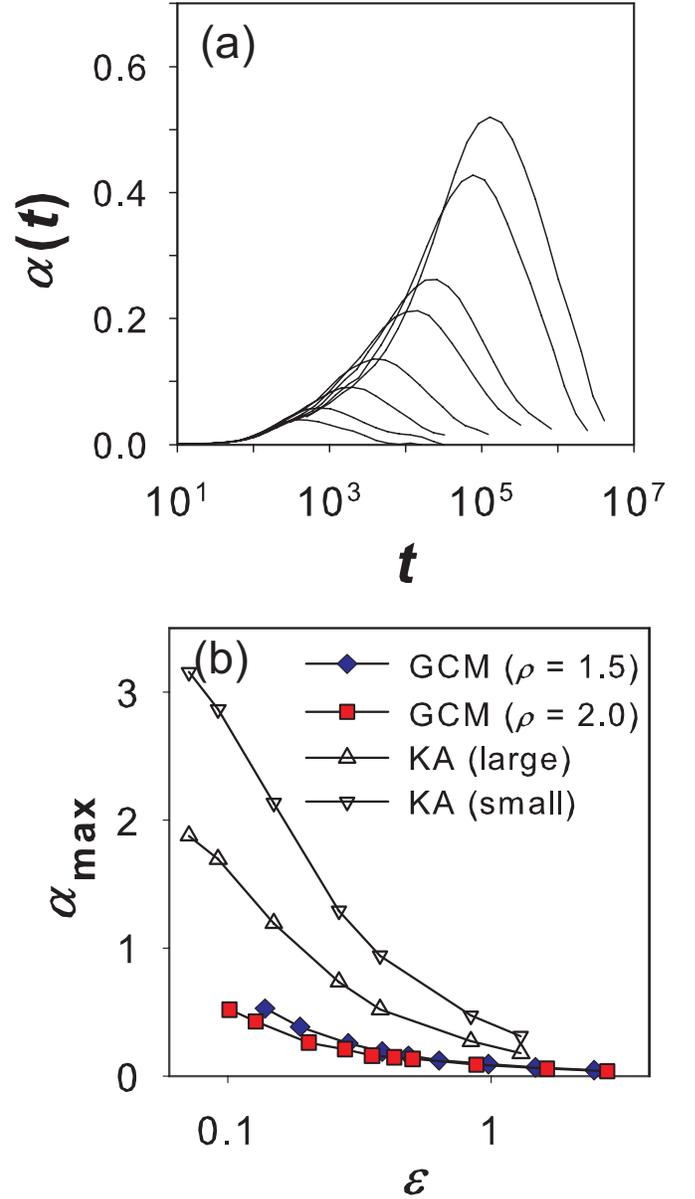}
\caption{
(a) The non-Gaussian parameter $\alpha(t)$ for $T\times 10^6  = 10$, 7, 5, 4, 3.4, 3.2 3 and 2.93 at $\rho=2.0$. 
(b) The temperature dependence of the maximum value of $\alpha(t)$ at $\rho=1.5$ (diamonds) and 2.0 (squares). 
The results for the large (up triangles) and small (down triangles)
 particles of the KA model are also plotted~\cite{Kob1994}. 
}
\vspace*{-0.1cm}
\label{alpha}
\end{center}
\end{figure}

Another good measure to monitor the extent of the departure from
the mean field behavior is the non-Gaussianity of the dynamics. 
At high temperatures, $F_s(k,t)$ or its real space expression, 
$G_s(r,t)\equiv \sum_i \ave{\delta(|\vec{R}_i(t) - \vec{R}_i(0)|-r)}$,
also known as the van Hove function, becomes  
almost a Gaussian function.
However, as the temperature is lowered to the supercooled regime, these
function substantially deviates from the Gaussian. 
This deviation is also considered to be a manifestation of dynamic heterogeneities.  
To quantify this, it is common to introduce the non-Gaussian parameter defined by 
\begin{equation}
\alpha(t) \equiv \frac{3\ave{R^4(t)}}{5\ave{R^2(t)}^2} -1,
\end{equation} 
where $\ave{R^4(t)} = N^{-1}\sum_i \ave{|\vec{R}_i(t) - \vec{R}_i(0)|^4}$. 
In Fig.~\ref{alpha} (a), we plot $\alpha(t)$ for $\rho=2.0$ at several temperatures.
It shows typical behaviors of the supercooled liquids, characterized by 
pronounced peaks at $t$ near or slightly before $\tau_{\alpha}$ whose heights increase
as the temperature decreases. 
However, the heights of the peaks are considerably lower than other model
glassformers at the comparable reduced temperatures $\varepsilon$. 
Fig.~\ref{alpha} (b) shows the temperature dependence of the maximum
value of the non-Gaussian parameter $\alpha_{max}$ for both $\rho=1.5$ and
2.0.   
The results for the KA model are also plotted~\cite{Kob1994}. 
Similarly to the result for the SE violation, $\alpha_{max}$ of the
GCM is far smaller than that of the KA model. 
Furthermore, one observes that $\alpha_{max}$ for $\rho=2.0$ is slightly smaller than that
for $\rho=1.5$. 
These results suggest that the dynamic heterogeneities are suppressed for the GCM and 
the suppression is stronger at  higher densities.
This is another collateral support that the high density GCM is 
more ``mean-field-like'' than other glassformers. 

\begin{figure}[t]
\begin{center}
\includegraphics[width=1\columnwidth]{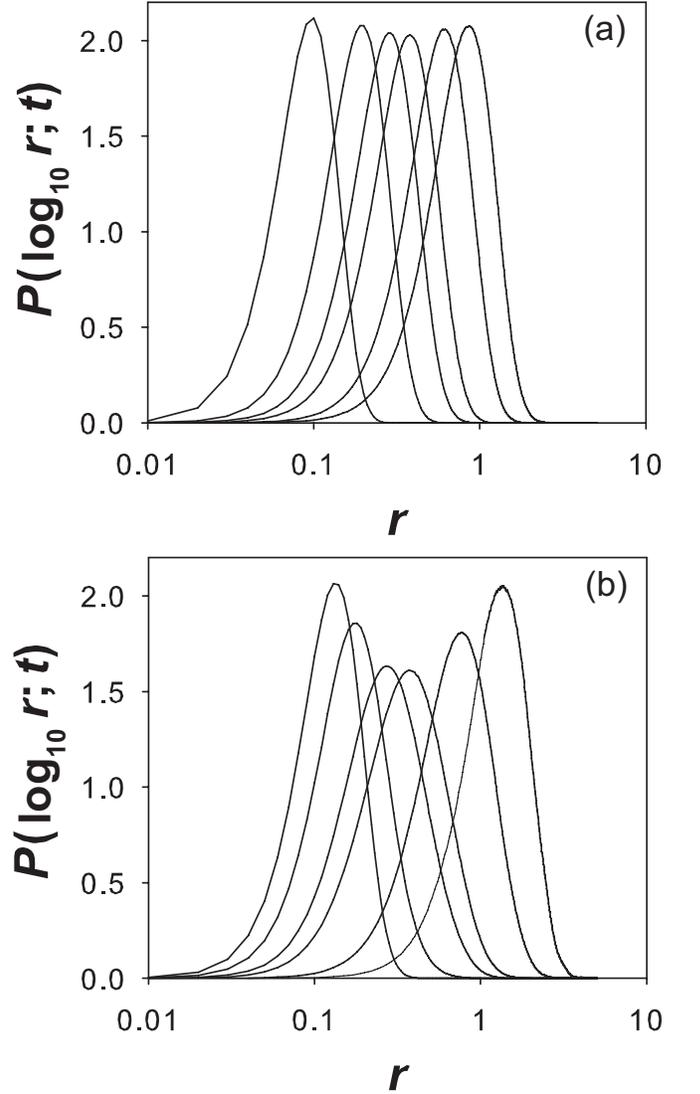}
\caption{
The probability distribution of the logarithm of the particle displacements. 
(a) The results for $T = 7.0 \times 10^{-6}$ and $\rho=2.0$. 
From left to right, $t$ = 44, 180, 512, 1024, 2896 and 5792.
(b) The results for $T = 2.93 \times 10^{-6}$ and $\rho=2.0$.
 From left to right, $t$ = 500, 32000, 181000, 362000, 1448000, and 4096000. 
}
\vspace*{-0.1cm}
\label{disp}
\end{center}
\end{figure}
More direct evidence that the dynamics of the high density GCM is closer
to a Gaussian process
and dynamic heterogeneities are weaker
can be obtained by monitoring the probability distribution of the
particle displacement $r$, denoted as $P(\log_{10} r;t)$. 
$P(\log_{10} r;t)$ is related to the van Hove function $G_s(r,t)$
by~\cite{Cates2004,Reichman2005,Flenner2005e}  
\begin{equation}
P(\log_{10} r;t)= (\ln 10) 4 \pi r^3 G_s(r,t).
\label{eq:Plog}
\end{equation}
If the dynamics is purely a Gaussian process,  $G_s(r,t)$ also becomes a Gaussian function, 
\begin{equation}
G_s(r,t) = \left( \frac{3}{2\pi\ave{R^2(t)}} \right)^{3/2} e^{-3r^2/2\ave{R^2(t)}}. 
\label{eq:GsGauss}
\end{equation}
From Eqs.~(\ref{eq:Plog}) and (\ref{eq:GsGauss}), $P(\log_{10} r;t)$ becomes a function of solely
$r/\sqrt{\ave{R^2(t)}}$; 
\begin{equation}
P(\log_{10} r;t) = (\ln 10) 4 \pi  \left( \frac{3 r^2}{2\pi\ave{R^2(t)}} \right)^{3/2} e^{-3r^2/2\ave{R^2(t)}}.
\label{eq:GsGauss2}
\end{equation}
Thus, the shape of $P(\log_{10} r;t)$ for a Gaussian process 
should be unchanged as $t$ is varied, 
but only shifted if plotted as a function of $\log_{10} r$.   
The peak height should be a constant value of $2.13$.  
In Fig.~\ref{disp}, 
we plotted the simulated $P(\log_{10} r;t)$ for  $\rho=2.0$ 
at the two temperatures;
$T= 7.0\times 10^{-6}$ ($\varepsilon \approx 1.2$)
and 
$T= 2.93\times 10^{-6}$ ($\varepsilon \approx 0.075$).
The high temperature result in Fig.~\ref{disp} (a) shows that 
$P(\log_{10} r;t)$ is almost given by Eq.~(\ref{eq:GsGauss2});
the shape of the function is almost Gaussian
and the peak height remains very close to $2.13$  over the long time. 
On the other hand, 
the low temperature result in Fig.~\ref{disp} (b)   shows that 
the peak height of the function becomes lower and the width becomes slightly larger 
at $t \sim \tau_{\alpha}$.  
This non-Gaussian behavior at the beta to alpha relaxation time regime 
is a common properties of $P(\log_{10} r;t)$ at a mildly supercooled state.
Note that, however, the extent of the non-Gaussianity shown in
Fig.~\ref{disp} (b) is much weaker than that of other glassformers such as the
KA model~\cite{Flenner2005e}. 
$P(\log_{10} r;t)$ for typical model glassformers
is known to split into the binodal shape at low temperatures, 
corresponding to a separation of the constituent particles 
into the mobile and immobile ones. 
This is one of the most salient feature of the dynamic heterogeneities. 
The peak of $P(\log_{10} r;t)$ in Fig.~\ref{disp} (b) does not show any
hint to split into the binodal shape. 
$P(\log_{10} r; t)$ of the KA model at $\varepsilon = 0.08$ ($T = 0.47$
in the LJ unit), 
a comparable reduced temperature as that of Fig.~\ref{disp} (b),  
is completely separated to the two peaks, corresponding to the
distribution of mobile and immobile particles.   
The decrease of the peak height of $P(\log_{10} r; t)$ in Fig.~\ref{disp} (b) 
is compatible with that of the KA model at much higher temperature, 
$\varepsilon = 0.38$ ($T = 0.6$ in the LJ unit)~\cite{Flenner2005e}.  
Above results strongly suggest that the dynamics of the high density GCM is
more Gaussian-like than that of the conventional model glassformers and the
dynamic heterogeneities are strongly suppressed.

\section{Summary and Outlook}

In this paper, we presented the detailed analysis of dynamics of the
high density GCM. 
The results are summarized below. 

(i) 
The crystal nucleation becomes slower as the density
increases. 
Analysis of the two orientational bond order parameters, $\bar{q}_4$ and
$\bar{q}_6$, reveals that the crystal structure is bcc at all densities
beyond the reentrant point.  

(ii) The system which failed to crystallize shows clear two-step and stretched exponential relaxation 
in the (both self and collective) intermediate scattering functions, which is the hallmarks of the
supercooled fluid near the glass transition point. 
All dynamical properties which we have analyzed are well
described by MCT. 
First, the temperature dependence of the diffusion coefficient and the
structural relaxation time is well fitted by the MCT power law. 
The parameter $T\simmct$ used to fit the simulation data is
unprecedentedly close to the theoretical prediction. 
The time dependence of the self intermediate scattering function $F_s(k,t)$
is well fitted by MCT, using the reduced temperature $\varepsilon$ as a sole parameter.
Furthermore, the nonergodic parameters for both collective and self
intermediate scattering functions, $F_{\infty}(k)$ and $F_{s,\infty}(k)$, 
are well described MCT. 
Here we find two noticeable differences from the typical glassformers. 
First, the shape of $F_{\infty}(k)$ is qualitatively different from
$F_{s,\infty}(k)$ at small wavevectors regime, where$F(k, t)$ decays very fast 
and the nonergodic parameter vanishes, whereas $F_s(k,t)$ decays very
slowly and its nonergodic parameter remains finite down to $k=0$.
We conjecture that this decoupling of the collective density dynamics
from the single particle dynamics is universal for the systems with the
long-ranged interactions. 
This indicates that the large-scale density fluctuation is decoupled to
the slow structural relaxation processes.  
Similar decoupling has been predicted from the MCT analysis of the systems with the power law
interactions $v(r) \sim 1/r^{n}$ with small $n$~\cite{Shiroiwa2010}.
Second, the agreement between MCT and simulation for $F_{s,\infty}(k)$ 
is satisfactory but conceivably worse than those for other model glassformers
such as the KA model~\cite{Gleim1998,Foffi2004}. 
We found a weak shoulder at the intermediate wavevectors.
This shoulder is reminiscent of those found in the MCT analysis of the
hard sphere glasses at the high dimensions~\cite{Ikeda2010}. 
We conjecture that the anomalous shoulders are the deficiency of MCT 
which appears only at the mean-field limit. 

(iii) Dynamic heterogeneities are suppressed in 
the high density GCM. 
The SE violation is very weak and the peak height of the non-Gaussian
parameter is much lower than the conventional model glassformers at the comparable
reduced temperatures. 
The weak dynamic heterogeneities of 
the high density  GCM was most obvious from
the observation of the probability distribution of the particle
displacement $P(\log_{10} r; t)$. 
We find no obvious change in the shape of 
$P(\log_{10} r; t)$ which remains almost Gaussian, though the width 
slightly widens around the beta to alpha relaxation time regimes.
Even at the lowest reduced temperature, at which the typical
model glassformers exhibit the very clear binodal distribution 
of mobile and immobile particles, due to the growing dynamic
heterogeneities, the probability distribution of the GCM remains to be a
single peak function. 

We conclude that the high density GCM is the ideal model system to
study the glass transition.
It is not only the cleanest glass model in that it is the one-component
system. 
But it is also the closest to the {\it ``mean-field''} model in that
dynamic heterogeneities are strongly suppressed
and the way how MCT predicts simulation results is 
synchronized with the way it does for the high dimensional systems.
The mean-field nature comes from the long-range nature of the
interaction potential, which is caused by the overlapping of the
particles at the high densities. 
Both the excellent agreement with MCT and small deviation from MCT (the
shoulder of $F_{s,\infty}(k)$) also lead us to reconsider the validity of
MCT as the the mean field theory of the glass transition. 
Mean-field models of the glass transition have been proposed and
analyzed by taking the long-range limit of the
interactions~\cite{Dotsenko2004} but 
it has never been realized in the simulation box. 
The another mean field limit, {\it i.e.}, the high dimension
limit, is another interesting challenge but given the current CPU power, 
going beyond $d=5$ would be unrealistic.
In this sense, 
the high density GCM might be the first realistic fluid model
which may be able to bridge the gap between the finite dimensional
system with the mean-field limit. 
It is tempting to consider the high density limit of the GCM where  
the small parameter $1/\rho$ may make the analytical
treatment of especially the static/thermodynamic parameters tractable and
leads us the {\it exact} mode-coupling theory (or alike). 

\acknowledgments
This work is partially supported by Grant-in-Aid for JSPS Fellows (AI),
KAKENHI; \# 21540416 (KM), and Priority Areas ``Soft Matter Physics''
(KM).


\begin{thebibliography}{79}
\expandafter\ifx\csname natexlab\endcsname\relax\def\natexlab#1{#1}\fi
\expandafter\ifx\csname bibnamefont\endcsname\relax
  \def\bibnamefont#1{#1}\fi
\expandafter\ifx\csname bibfnamefont\endcsname\relax
  \def\bibfnamefont#1{#1}\fi
\expandafter\ifx\csname citenamefont\endcsname\relax
  \def\citenamefont#1{#1}\fi
\expandafter\ifx\csname url\endcsname\relax
  \def\url#1{\texttt{#1}}\fi
\expandafter\ifx\csname urlprefix\endcsname\relax\def\urlprefix{URL }\fi
\providecommand{\bibinfo}[2]{#2}
\providecommand{\eprint}[2][]{\url{#2}}

\bibitem[{\citenamefont{Debenedetti and Stillinger}((2001))}]{Debenedetti2001}
\bibinfo{author}{\bibfnamefont{P.~G.} \bibnamefont{Debenedetti}}
  \bibnamefont{and} \bibinfo{author}{\bibfnamefont{F.~H.}
  \bibnamefont{Stillinger}}, \bibinfo{journal}{Nature}
  \textbf{\bibinfo{volume}{{\bf 410}}}, \bibinfo{pages}{259}
  (\bibinfo{year}{2001}).

\bibitem[{\citenamefont{Cavagna}(())}]{Cavagna2009b}
\bibinfo{author}{\bibfnamefont{A.}~\bibnamefont{Cavagna}}, 
\bibinfo{journal}{Phys. Rep.}
  \textbf{\bibinfo{volume}{{\bf 476}}}, \bibinfo{pages}{51}
  (\bibinfo{year}{2009}).

\bibitem[{\citenamefont{Biroli and Bouchaud}(())}]{Biroli2009}
\bibinfo{author}{\bibfnamefont{G.}~\bibnamefont{Biroli}} \bibnamefont{and}
  \bibinfo{author}{\bibfnamefont{J.~P.} \bibnamefont{Bouchaud}},
  \bibinfo{pages}{arXiv:0912.2542}.

\bibitem[{\citenamefont{Berthier et~al.}((2011))\citenamefont{Berthier, Biroli,
  Bouchaud, Cipelletti, and {van Saarloos}}}]{Berthier2011d}
\bibinfo{editor}{\bibfnamefont{L.}~\bibnamefont{Berthier}},
  \bibinfo{editor}{\bibfnamefont{G.}~\bibnamefont{Biroli}},
  \bibinfo{editor}{\bibfnamefont{J.-P.} \bibnamefont{Bouchaud}},
  \bibinfo{editor}{\bibfnamefont{L.}~\bibnamefont{Cipelletti}},
  \bibnamefont{and} \bibinfo{editor}{\bibfnamefont{W.}~\bibnamefont{{van
  Saarloos}}}, eds., \emph{\bibinfo{title}{{Dynamical Heterogeneities in
  Glasses, Colloids, and Granular Media}}} (\bibinfo{publisher}{Oxford
  University Press, Oxford}, \bibinfo{year}{2011}).

\bibitem[{\citenamefont{Ikeda and Miyazaki}((2010))}]{Ikeda2010}
\bibinfo{author}{\bibfnamefont{A.}~\bibnamefont{Ikeda}} \bibnamefont{and}
  \bibinfo{author}{\bibfnamefont{K.}~\bibnamefont{Miyazaki}},
  \bibinfo{journal}{Phys. Rev. Lett.} \textbf{\bibinfo{volume}{{\bf 104}}},
  \bibinfo{pages}{255704} (\bibinfo{year}{2010});
  \bibinfo{journal}{{\it ibid.}} \textbf{\bibinfo{volume}{{\bf 106}}},
  \bibinfo{pages}{049602} (\bibinfo{year}{2011}{\natexlab{b}}).

\bibitem[{\citenamefont{Schmid and Schilling}((2010))}]{Schmid2010b}
\bibinfo{author}{\bibfnamefont{B.}~\bibnamefont{Schmid}} \bibnamefont{and}
  \bibinfo{author}{\bibfnamefont{R.}~\bibnamefont{Schilling}},
  \bibinfo{journal}{Phys. Rev. {\rm E}} \textbf{\bibinfo{volume}{{\bf 81}}},
  \bibinfo{pages}{041502} (\bibinfo{year}{2010});
\bibinfo{author}{\bibfnamefont{R.}~\bibnamefont{Schilling}} \bibnamefont{and}
  \bibinfo{author}{\bibfnamefont{B.}~\bibnamefont{Schmid}},
  \bibinfo{journal}{Phys. Rev. Lett.} \textbf{\bibinfo{volume}{{\bf 106}}},
  \bibinfo{pages}{049601} (\bibinfo{year}{2011}).

\bibitem[{\citenamefont{Andersen}((2005))}]{Andersen2005}
\bibinfo{author}{\bibfnamefont{H.~C.} \bibnamefont{Andersen}},
  \bibinfo{journal}{Proc. Natl. Acad. Sci. U. S. A.}
  \textbf{\bibinfo{volume}{{\bf 102}}}, \bibinfo{pages}{6686}
  (\bibinfo{year}{2005}).

\bibitem[{\citenamefont{Berthier and Tarjus}((2009))}]{Berthier2009d}
\bibinfo{author}{\bibfnamefont{L.}~\bibnamefont{Berthier}} \bibnamefont{and}
  \bibinfo{author}{\bibfnamefont{G.}~\bibnamefont{Tarjus}},
  \bibinfo{journal}{Phys. Rev. Lett.} \textbf{\bibinfo{volume}{{\bf 103}}},
  \bibinfo{pages}{170601} (\bibinfo{year}{2009}).

\bibitem[{\citenamefont{Sausset and Tarjus}((2010))}]{Sausset2010b}
\bibinfo{author}{\bibfnamefont{F.}~\bibnamefont{Sausset}} \bibnamefont{and}
  \bibinfo{author}{\bibfnamefont{G.}~\bibnamefont{Tarjus}},
  \bibinfo{journal}{Phys. Rev. Lett.} \textbf{\bibinfo{volume}{{\bf 104}}},
  \bibinfo{pages}{065701} (\bibinfo{year}{2010}).

\bibitem[{\citenamefont{Charbonneau et~al.}((2010))\citenamefont{Charbonneau,
  Ikeda, {van Meel}, and Miyazaki}}]{Charbonneau2010}
\bibinfo{author}{\bibfnamefont{P.}~\bibnamefont{Charbonneau}},
  \bibinfo{author}{\bibfnamefont{A.}~\bibnamefont{Ikeda}},
  \bibinfo{author}{\bibfnamefont{J.~A.} \bibnamefont{{van Meel}}},
  \bibnamefont{and} \bibinfo{author}{\bibfnamefont{K.}~\bibnamefont{Miyazaki}},
  \bibinfo{journal}{Phys. Rev. {\rm E}} \textbf{\bibinfo{volume}{{\bf 81}}},
  \bibinfo{pages}{040501(R)} (\bibinfo{year}{2010}).

\bibitem[{\citenamefont{Kirkpatrick et~al.}((1989))\citenamefont{Kirkpatrick,
  Thirumalai, and Wolynes}}]{Kirkpatrick1989}
\bibinfo{author}{\bibfnamefont{T.~R.} \bibnamefont{Kirkpatrick}},
  \bibinfo{author}{\bibfnamefont{D.}~\bibnamefont{Thirumalai}},
  \bibnamefont{and} \bibinfo{author}{\bibfnamefont{P.~G.}
  \bibnamefont{Wolynes}}, \bibinfo{journal}{Phys. Rev. {\rm A}}
  \textbf{\bibinfo{volume}{{\bf 40}}}, \bibinfo{pages}{1045}
  (\bibinfo{year}{1989}).

\bibitem[{\citenamefont{{M\'{e}zard} and Parisi}((1999))}]{Mezard1999}
\bibinfo{author}{\bibfnamefont{M.}~\bibnamefont{{M\'{e}zard}}}
  \bibnamefont{and} \bibinfo{author}{\bibfnamefont{G.}~\bibnamefont{Parisi}},
  \bibinfo{journal}{Phys. Rev. Lett.} \textbf{\bibinfo{volume}{{\bf 82}}},
  \bibinfo{pages}{747} (\bibinfo{year}{1999}).

\bibitem[{\citenamefont{Parisi and Zamponi}((2010))}]{Parisi2010}
\bibinfo{author}{\bibfnamefont{G.}~\bibnamefont{Parisi}} \bibnamefont{and}
  \bibinfo{author}{\bibfnamefont{F.}~\bibnamefont{Zamponi}},
  \bibinfo{journal}{Rev. Mod. Phys.} \textbf{\bibinfo{volume}{{\bf 82}}},
  \bibinfo{pages}{789} (\bibinfo{year}{2010}).

\bibitem[{\citenamefont{G{\"{o}}tze}((2009))}]{Gotze2009}
\bibinfo{author}{\bibfnamefont{W.}~\bibnamefont{G{\"{o}}tze}},
  \emph{\bibinfo{title}{{"Complex Dynamics of Glass-Forming Liquids"}}}
  (\bibinfo{publisher}{Oxford University Press, Oxford},
  \bibinfo{year}{2009}).

\bibitem[{\citenamefont{Lubchenko and Wolynes}((2007))}]{Lubchenko2007}
\bibinfo{author}{\bibfnamefont{V.}~\bibnamefont{Lubchenko}} \bibnamefont{and}
  \bibinfo{author}{\bibfnamefont{P.~G.} \bibnamefont{Wolynes}},
  \bibinfo{journal}{Annu. Rev. Phys. Chem.} \textbf{\bibinfo{volume}{{\bf
  58}}}, \bibinfo{pages}{235} (\bibinfo{year}{2007}).

\bibitem[{\citenamefont{Eaves and Reichman}((2009))}]{Eaves2009}
\bibinfo{author}{\bibfnamefont{J.~D.} \bibnamefont{Eaves}} \bibnamefont{and}
  \bibinfo{author}{\bibfnamefont{D.~R.} \bibnamefont{Reichman}},
  \bibinfo{journal}{Proc. Natl. Acad. Sci. U. S. A.}
  \textbf{\bibinfo{volume}{{\bf 106}}}, \bibinfo{pages}{15171}
  (\bibinfo{year}{2009}).

\bibitem[{\citenamefont{Biroli and Bouchaud}((2007))}]{Biroli2007b}
\bibinfo{author}{\bibfnamefont{G.}~\bibnamefont{Biroli}} \bibnamefont{and}
  \bibinfo{author}{\bibfnamefont{J.-P.} \bibnamefont{Bouchaud}},
  \bibinfo{journal}{J. Phys.: Condens. Matter} \textbf{\bibinfo{volume}{{\bf
  19}}}, \bibinfo{pages}{205101} (\bibinfo{year}{2007}).

\bibitem[{\citenamefont{Biroli et~al.}((2006))\citenamefont{Biroli, Bouchaud,
  Miyazaki, and Reichman}}]{Biroli2006b}
\bibinfo{author}{\bibfnamefont{G.}~\bibnamefont{Biroli}},
  \bibinfo{author}{\bibfnamefont{J.-P.} \bibnamefont{Bouchaud}},
  \bibinfo{author}{\bibfnamefont{K.}~\bibnamefont{Miyazaki}}, \bibnamefont{and}
  \bibinfo{author}{\bibfnamefont{D.~R.} \bibnamefont{Reichman}},
  \bibinfo{journal}{Phys. Rev. Lett.} \textbf{\bibinfo{volume}{{\bf 97}}},
  \bibinfo{pages}{195701} (\bibinfo{year}{2006}).

\bibitem[{\citenamefont{Zaccarelli et~al.}((2008))\citenamefont{Zaccarelli,
  Andreev, and Reichman}}]{Zaccarelli2008b}
\bibinfo{author}{\bibfnamefont{E.}~\bibnamefont{Zaccarelli}},
  \bibinfo{author}{\bibfnamefont{F.}~\bibnamefont{Andreev},
  \bibfnamefont{Stefan~Sciortino}}, \bibnamefont{and}
  \bibinfo{author}{\bibfnamefont{D.~R.} \bibnamefont{Reichman}},
  \bibinfo{journal}{Phys. Rev. Lett.} \textbf{\bibinfo{volume}{{\bf 100}}},
  \bibinfo{pages}{195701} (\bibinfo{year}{2008}).

\bibitem[{\citenamefont{Dotsenko}((2004))}]{Dotsenko2004}
\bibinfo{author}{\bibfnamefont{V.~S.}~\bibnamefont{Dotsenko}},
  \bibinfo{journal}{J. Stat. Phys.} \textbf{\bibinfo{volume}{{\bf 115}}},
  \bibinfo{pages}{823} (\bibinfo{year}{2004});
\bibinfo{author}{\bibfnamefont{V.~S.} \bibnamefont{Dotsenko}} \bibnamefont{and}
  \bibinfo{author}{\bibfnamefont{G.}~\bibnamefont{Blatter}},
  \bibinfo{journal}{Phys. Rev. {\rm E}} \textbf{\bibinfo{volume}{{\bf 72}}},
  \bibinfo{pages}{021502} (\bibinfo{year}{2005}).

\bibitem[{\citenamefont{Mari and Kurchan}(())}]{Mari2011}
\bibinfo{author}{\bibfnamefont{R.}~\bibnamefont{Mari}} \bibnamefont{and}
  \bibinfo{author}{\bibfnamefont{J.}~\bibnamefont{Kurchan}},
  \bibinfo{pages}{arXiv:1104.3420}.

\bibitem[{\citenamefont{Stillinger}((1976))}]{Stillinger1976}
\bibinfo{author}{\bibfnamefont{F.~H.} \bibnamefont{Stillinger}},
  \bibinfo{journal}{J. Chem. Phys.} \textbf{\bibinfo{volume}{{\bf 65}}},
  \bibinfo{pages}{3968} (\bibinfo{year}{1976});
\bibinfo{author}{\bibfnamefont{F.~H.} \bibnamefont{Stillinger}}
  \bibnamefont{and} \bibinfo{author}{\bibfnamefont{T.~A.} \bibnamefont{Weber}},
  \bibinfo{journal}{{\it ibid.}} \textbf{\bibinfo{volume}{{\bf 68}}},
  \bibinfo{pages}{3837} (\bibinfo{year}{1978});
\bibinfo{author}{\bibfnamefont{F.~H.} \bibnamefont{Stillinger}}
  \bibnamefont{and} \bibinfo{author}{\bibfnamefont{T.~A.} \bibnamefont{Weber}},
  \bibinfo{journal}{{\it ibid.}} \textbf{\bibinfo{volume}{{\bf 70}}},
  \bibinfo{pages}{4879} (\bibinfo{year}{1979});
\bibinfo{author}{\bibfnamefont{F.~H.} \bibnamefont{Stillinger}},
  \bibinfo{journal}{Phys. Rev. {\rm B}} \textbf{\bibinfo{volume}{{\bf 20}}},
  \bibinfo{pages}{299} (\bibinfo{year}{1979}).

\bibitem[{\citenamefont{Stillinger and Stillinger}((1997))}]{Stillinger1997}
\bibinfo{author}{\bibfnamefont{F.~H.} \bibnamefont{Stillinger}}
  \bibnamefont{and} \bibinfo{author}{\bibfnamefont{D.~K.}
  \bibnamefont{Stillinger}}, \bibinfo{journal}{Physica {\rm A}}
  \textbf{\bibinfo{volume}{{\bf 244}}}, \bibinfo{pages}{358}
  (\bibinfo{year}{1997}).

\bibitem[{\citenamefont{Lang et~al.}((2000))\citenamefont{Lang, Likos,
  Watzlawek, and L{\"{o}}wen}}]{Lang2000}
\bibinfo{author}{\bibfnamefont{A.}~\bibnamefont{Lang}},
  \bibinfo{author}{\bibfnamefont{C.~N.} \bibnamefont{Likos}},
  \bibinfo{author}{\bibfnamefont{M.}~\bibnamefont{Watzlawek}},
  \bibnamefont{and}
  \bibinfo{author}{\bibfnamefont{H.}~\bibnamefont{L{\"{o}}wen}},
  \bibinfo{journal}{J. Phys.: Condens. Matter} \textbf{\bibinfo{volume}{{\bf
  12}}}, \bibinfo{pages}{5087} (\bibinfo{year}{2000}).

\bibitem[{\citenamefont{Louis et~al.}((2000))\citenamefont{Louis, Bolhuis, and
  Hansen}}]{Louis2000b}
\bibinfo{author}{\bibfnamefont{A.~A.} \bibnamefont{Louis}},
  \bibinfo{author}{\bibfnamefont{P.~G.} \bibnamefont{Bolhuis}},
  \bibnamefont{and} \bibinfo{author}{\bibfnamefont{J.~P.}
  \bibnamefont{Hansen}}, \bibinfo{journal}{Phys. Rev. {\rm E}}
  \textbf{\bibinfo{volume}{{\bf 62}}}, \bibinfo{pages}{7961}
  (\bibinfo{year}{2000}).

\bibitem[{\citenamefont{Prestipino
  et~al.}((2005){\natexlab{a}})\citenamefont{Prestipino, Saija, and
  Giaquinta}}]{Prestipino2005}
\bibinfo{author}{\bibfnamefont{S.}~\bibnamefont{Prestipino}},
  \bibinfo{author}{\bibfnamefont{F.}~\bibnamefont{Saija}}, \bibnamefont{and}
  \bibinfo{author}{\bibfnamefont{P.~V.} \bibnamefont{Giaquinta}},
  \bibinfo{journal}{Phys. Rev. {\rm E}} \textbf{\bibinfo{volume}{{\bf 71}}},
  \bibinfo{pages}{050102(R)} (\bibinfo{year}{2005}{\natexlab{a}});
\bibinfo{author}{\bibfnamefont{S.}~\bibnamefont{Prestipino}},
  \bibinfo{author}{\bibfnamefont{F.}~\bibnamefont{Saija}}, \bibnamefont{and}
  \bibinfo{author}{\bibfnamefont{P.~V.} \bibnamefont{Giaquinta}},
  \bibinfo{journal}{J. Chem. Phys.} \textbf{\bibinfo{volume}{{\bf 123}}},
  \bibinfo{pages}{144110} (\bibinfo{year}{2005}{\natexlab{b}}).

\bibitem[{\citenamefont{Mladek et~al.}((2006))\citenamefont{Mladek, Gottwald,
  Kahl, Neumann, and Likos}}]{Mladek2006}
\bibinfo{author}{\bibfnamefont{B.~M.} \bibnamefont{Mladek}},
  \bibinfo{author}{\bibfnamefont{D.}~\bibnamefont{Gottwald}},
  \bibinfo{author}{\bibfnamefont{G.}~\bibnamefont{Kahl}},
  \bibinfo{author}{\bibfnamefont{M.}~\bibnamefont{Neumann}}, \bibnamefont{and}
  \bibinfo{author}{\bibfnamefont{C.~N.} \bibnamefont{Likos}},
  \bibinfo{journal}{Phys. Rev. Lett.} \textbf{\bibinfo{volume}{{\bf 96}}},
  \bibinfo{pages}{045701} (\bibinfo{year}{2006}).

\bibitem[{\citenamefont{Mausbach and May}((2006))}]{Mausbach2006}
\bibinfo{author}{\bibfnamefont{P.}~\bibnamefont{Mausbach}} \bibnamefont{and}
  \bibinfo{author}{\bibfnamefont{H.-O.} \bibnamefont{May}},
  \bibinfo{journal}{Fluid Phase Equilibria} \textbf{\bibinfo{volume}{{\bf
  249}}}, \bibinfo{pages}{17} (\bibinfo{year}{2006}).

\bibitem[{\citenamefont{Zachary et~al.}((2008))\citenamefont{Zachary,
  Stillinger, and Torquato}}]{Zachary2008}
\bibinfo{author}{\bibfnamefont{C.~E.} \bibnamefont{Zachary}},
  \bibinfo{author}{\bibfnamefont{F.~H.} \bibnamefont{Stillinger}},
  \bibnamefont{and} \bibinfo{author}{\bibfnamefont{S.}~\bibnamefont{Torquato}},
  \bibinfo{journal}{J. Chem. Phys.} \textbf{\bibinfo{volume}{{\bf 128}}},
  \bibinfo{pages}{224505} (\bibinfo{year}{2008}).

\bibitem[{\citenamefont{Krekelberg
  et~al.}((2009){\natexlab{a}})\citenamefont{Krekelberg, Kumar, Mittal,
  Errington, and Truskett}}]{Krekelberg2009c}
\bibinfo{author}{\bibfnamefont{W.~P.} \bibnamefont{Krekelberg}},
  \bibinfo{author}{\bibfnamefont{T.}~\bibnamefont{Kumar}},
  \bibinfo{author}{\bibfnamefont{J.}~\bibnamefont{Mittal}},
  \bibinfo{author}{\bibfnamefont{J.~R.} \bibnamefont{Errington}},
  \bibnamefont{and} \bibinfo{author}{\bibfnamefont{T.~M.}
  \bibnamefont{Truskett}}, \bibinfo{journal}{Phys. Rev. {\rm E}}
  \textbf{\bibinfo{volume}{{\bf 79}}}, \bibinfo{pages}{031203}
  (\bibinfo{year}{2009}{\natexlab{a}});
\bibinfo{author}{\bibfnamefont{W.~P.} \bibnamefont{Krekelberg}},
  \bibinfo{author}{\bibfnamefont{M.~J.} \bibnamefont{Pond}},
  \bibinfo{author}{\bibfnamefont{G.}~\bibnamefont{Goel}},
  \bibinfo{author}{\bibfnamefont{V.~K.} \bibnamefont{Shen}},
  \bibinfo{author}{\bibfnamefont{J.~R.} \bibnamefont{Errington}},
  \bibnamefont{and} \bibinfo{author}{\bibfnamefont{T.~M.}
  \bibnamefont{Truskett}}, \bibinfo{journal}{{\it ibid.}}
  \textbf{\bibinfo{volume}{{\bf 80}}}, \bibinfo{pages}{061205}
  (\bibinfo{year}{2009}{\natexlab{b}});
\bibinfo{author}{\bibfnamefont{M.~J.} \bibnamefont{Pond}},
  \bibinfo{author}{\bibfnamefont{W.~P.} \bibnamefont{Krekelberg}},
  \bibinfo{author}{\bibfnamefont{V.~K.} \bibnamefont{Shen}},
  \bibinfo{author}{\bibfnamefont{J.~R.} \bibnamefont{Errington}},
  \bibnamefont{and} \bibinfo{author}{\bibfnamefont{T.~M.}
  \bibnamefont{Truskett}}, \bibinfo{journal}{J. Chem. Phys.}
  \textbf{\bibinfo{volume}{{\bf 131}}}, \bibinfo{pages}{161101}
  (\bibinfo{year}{2009});\bibinfo{author}{\bibfnamefont{M.~J.} \bibnamefont{Pond}},
  \bibinfo{author}{\bibfnamefont{J.~R.} \bibnamefont{Errington}},
  \bibnamefont{and} \bibinfo{author}{\bibfnamefont{T.~M.}
  \bibnamefont{Truskett}}, \bibinfo{pages}{arXiv:1101.1982}.



\bibitem[{\citenamefont{Shall and Egorov}(())}]{Shall2010}
\bibinfo{author}{\bibfnamefont{L.~A.} \bibnamefont{Shall}} \bibnamefont{and}
  \bibinfo{author}{\bibfnamefont{S.~A.} \bibnamefont{Egorov}},
  \bibinfo{journal}{J. Chem. Phys.}
  \textbf{\bibinfo{volume}{{\bf 132}}}, \bibinfo{pages}{184504131}
  (\bibinfo{year}{2010}).

\bibitem[{\citenamefont{Likos}((2001))}]{Likos2001}
\bibinfo{author}{\bibfnamefont{C.~N.} \bibnamefont{Likos}},
  \bibinfo{journal}{Phys. Rep.} \textbf{\bibinfo{volume}{{\bf 348}}},
  \bibinfo{pages}{267} (\bibinfo{year}{2001});
  \bibinfo{journal}{Soft Matter} \textbf{\bibinfo{volume}{{\bf 2}}},
  \bibinfo{pages}{478} (\bibinfo{year}{2006}).

\bibitem[{\citenamefont{Ikeda and Miyazaki}((2011){\natexlab{a}})}]{Ikeda2011}
\bibinfo{author}{\bibfnamefont{A.}~\bibnamefont{Ikeda}} \bibnamefont{and}
  \bibinfo{author}{\bibfnamefont{K.}~\bibnamefont{Miyazaki}},
  \bibinfo{journal}{Phys. Rev. Lett.} \textbf{\bibinfo{volume}{{\bf 106}}},
  \bibinfo{pages}{015701} (\bibinfo{year}{2011}{\natexlab{a}}).

\bibitem[{\citenamefont{Ikeda and Miyazaki}((unpublished))}]{Ikeda_I}
\bibinfo{author}{\bibfnamefont{A.}~\bibnamefont{Ikeda}} \bibnamefont{and}
  \bibinfo{author}{\bibfnamefont{K.}~\bibnamefont{Miyazaki}},
  \textbf{\bibinfo{volume}{{\bf }}} (\bibinfo{year}{unpublished}).

\bibitem[{\citenamefont{Foffi et~al.}((2003))\citenamefont{Foffi, Sciortino,
  Tartaglia, Zaccarelli, Verso, Reatto, Dawson, and Likos}}]{Foffi2003b}
\bibinfo{author}{\bibfnamefont{G.}~\bibnamefont{Foffi}},
  \bibinfo{author}{\bibfnamefont{F.}~\bibnamefont{Sciortino}},
  \bibinfo{author}{\bibfnamefont{P.}~\bibnamefont{Tartaglia}},
  \bibinfo{author}{\bibfnamefont{E.}~\bibnamefont{Zaccarelli}},
  \bibinfo{author}{\bibfnamefont{F.~L.} \bibnamefont{Verso}},
  \bibinfo{author}{\bibfnamefont{L.}~\bibnamefont{Reatto}},
  \bibinfo{author}{\bibfnamefont{K.~A.} \bibnamefont{Dawson}},
  \bibnamefont{and} \bibinfo{author}{\bibfnamefont{C.~N.} \bibnamefont{Likos}},
  \bibinfo{journal}{Phys. Rev. Lett.} \textbf{\bibinfo{volume}{{\bf 90}}},
  \bibinfo{pages}{238301} (\bibinfo{year}{2003}).

\bibitem[{\citenamefont{Zaccarelli et~al.}((2005))\citenamefont{Zaccarelli,
  Mayer, Asteriadi, Likos, Sciortino, Roovers, Iatrou, Hadjichristidis,
  Tartaglia, L{\"{o}}wen et~al.}}]{Zaccarelli2005c}
\bibinfo{author}{\bibfnamefont{E.}~\bibnamefont{Zaccarelli}},
  \bibinfo{author}{\bibfnamefont{C.}~\bibnamefont{Mayer}},
  \bibinfo{author}{\bibfnamefont{A.}~\bibnamefont{Asteriadi}},
  \bibinfo{author}{\bibfnamefont{C.~N.} \bibnamefont{Likos}},
  \bibinfo{author}{\bibfnamefont{F.}~\bibnamefont{Sciortino}},
  \bibinfo{author}{\bibfnamefont{J.}~\bibnamefont{Roovers}},
  \bibinfo{author}{\bibfnamefont{H.}~\bibnamefont{Iatrou}},
  \bibinfo{author}{\bibfnamefont{N.}~\bibnamefont{Hadjichristidis}},
  \bibinfo{author}{\bibfnamefont{P.}~\bibnamefont{Tartaglia}},
  \bibinfo{author}{\bibfnamefont{H.}~\bibnamefont{L{\"{o}}wen}},
  \bibnamefont{et~al.}, \bibinfo{journal}{Phys. Rev. Lett.}
  \textbf{\bibinfo{volume}{{\bf 95}}}, \bibinfo{pages}{268301}
  (\bibinfo{year}{2005});
\bibinfo{author}{\bibfnamefont{C.}~\bibnamefont{Mayer}},
  \bibinfo{author}{\bibfnamefont{E.}~\bibnamefont{Zaccarelli}},
  \bibinfo{author}{\bibfnamefont{E.}~\bibnamefont{Stiakakis}},
  \bibinfo{author}{\bibfnamefont{C.~N.} \bibnamefont{Likos}},
  \bibinfo{author}{\bibfnamefont{F.}~\bibnamefont{Sciortino}},
  \bibinfo{author}{\bibfnamefont{A.}~\bibnamefont{Munam}},
  \bibinfo{author}{\bibfnamefont{M.}~\bibnamefont{Gauthier}},
  \bibinfo{author}{\bibfnamefont{N.}~\bibnamefont{Hadjichristidis}},
  \bibinfo{author}{\bibfnamefont{H.}~\bibnamefont{Iatrou}},
  \bibinfo{author}{\bibfnamefont{P.}~\bibnamefont{Tartaglia}},
  \bibnamefont{et~al.}, \bibinfo{journal}{Nature Materials}
  \textbf{\bibinfo{volume}{{\bf 7}}}, \bibinfo{pages}{780}
  (\bibinfo{year}{2008}).

\bibitem[{\citenamefont{Berthier and
  Witten}((2009){\natexlab{a}})}]{Berthier2009c}
\bibinfo{author}{\bibfnamefont{L.}~\bibnamefont{Berthier}} \bibnamefont{and}
  \bibinfo{author}{\bibfnamefont{T.~A.} \bibnamefont{Witten}},
  \bibinfo{journal}{Europhys. Lett.} \textbf{\bibinfo{volume}{{\bf 86}}},
  \bibinfo{pages}{10001} (\bibinfo{year}{2009}{\natexlab{a}});
\bibinfo{author}{\bibfnamefont{L.}~\bibnamefont{Berthier}} \bibnamefont{and}
  \bibinfo{author}{\bibfnamefont{T.~A.} \bibnamefont{Witten}},
  \bibinfo{journal}{Phys. Rev. {\rm E}} \textbf{\bibinfo{volume}{{\bf 80}}},
  \bibinfo{pages}{021502} (\bibinfo{year}{2009}{\natexlab{b}}).

\bibitem[{\citenamefont{Berthier et~al.}((2010))\citenamefont{Berthier, Moreno,
  and Szamel}}]{Berthier2010i}
\bibinfo{author}{\bibfnamefont{L.}~\bibnamefont{Berthier}},
  \bibinfo{author}{\bibfnamefont{A.~J.} \bibnamefont{Moreno}},
  \bibnamefont{and} \bibinfo{author}{\bibfnamefont{G.}~\bibnamefont{Szamel}},
  \bibinfo{journal}{Phys. Rev. {\rm E}} \textbf{\bibinfo{volume}{{\bf 82}}},
  \bibinfo{pages}{060501(R)} (\bibinfo{year}{2010}).


\bibitem[{\citenamefont{Frenkel and Smit}((2001))}]{Frenkel2001}
\bibinfo{author}{\bibfnamefont{D.}~\bibnamefont{Frenkel}} \bibnamefont{and}
  \bibinfo{author}{\bibfnamefont{B.}~\bibnamefont{Smit}},
  \emph{\bibinfo{title}{{"Understanding Molecular Simulation"}}} (\bibinfo{publisher}{Academic Press},
  \bibinfo{year}{2001}).

\bibitem[{\citenamefont{Voigtmann}(())}]{Voigtmann2010}
\bibinfo{author}{\bibfnamefont{T.}~\bibnamefont{Voigtmann}},
  \bibinfo{pages}{arXiv:1010.0440}.

\bibitem[{\citenamefont{Binder and Kob}((2005))}]{Binder2005}
\bibinfo{author}{\bibfnamefont{K.}~\bibnamefont{Binder}} \bibnamefont{and}
  \bibinfo{author}{\bibfnamefont{W.}~\bibnamefont{Kob}},
  \emph{\bibinfo{title}{{"Glassy Materials and Disordered Solids"}}}
  (\bibinfo{publisher}{World Scientific, Singapore}, \bibinfo{year}{2005}).

\bibitem[{\citenamefont{Steinhardt et~al.}((1983))\citenamefont{Steinhardt,
  Nelson, and Ronchetti}}]{Steinhardt1983}
\bibinfo{author}{\bibfnamefont{P.~J.} \bibnamefont{Steinhardt}},
  \bibinfo{author}{\bibfnamefont{D.~R.} \bibnamefont{Nelson}},
  \bibnamefont{and}
  \bibinfo{author}{\bibfnamefont{M.}~\bibnamefont{Ronchetti}},
  \bibinfo{journal}{Phys. Rev. {\rm B}} \textbf{\bibinfo{volume}{{\bf 28}}},
  \bibinfo{pages}{784} (\bibinfo{year}{1983}).

\bibitem[{\citenamefont{Lechner and Dellago}((2008))}]{Lechner2008}
\bibinfo{author}{\bibfnamefont{W.}~\bibnamefont{Lechner}} \bibnamefont{and}
  \bibinfo{author}{\bibfnamefont{C.}~\bibnamefont{Dellago}},
  \bibinfo{journal}{J. Chem. Phys.} \textbf{\bibinfo{volume}{{\bf 129}}},
  \bibinfo{pages}{114707} (\bibinfo{year}{2008}).

\bibitem[{\citenamefont{Kawasaki and Tanaka}((2010))}]{Kawasaki2010}
\bibinfo{author}{\bibfnamefont{T.}~\bibnamefont{Kawasaki}} \bibnamefont{and}
  \bibinfo{author}{\bibfnamefont{H.}~\bibnamefont{Tanaka}},
  \bibinfo{journal}{Proc. Natl. Acad. Sci. U. S. A.}
  \textbf{\bibinfo{volume}{{\bf 107}}}, \bibinfo{pages}{14036}
  (\bibinfo{year}{2010}).

\bibitem[{\citenamefont{Ichimaru}((1982))}]{Ichimaru1982}
\bibinfo{author}{\bibfnamefont{S.}~\bibnamefont{Ichimaru}},
  \bibinfo{journal}{Rev. Mod. Phys.} \textbf{\bibinfo{volume}{{\bf 54}}},
  \bibinfo{pages}{1017} (\bibinfo{year}{1982}).

\bibitem[{\citenamefont{Kob and Andersen}((1994))}]{Kob1994}
\bibinfo{author}{\bibfnamefont{W.}~\bibnamefont{Kob}} \bibnamefont{and}
  \bibinfo{author}{\bibfnamefont{H.~C.} \bibnamefont{Andersen}},
  \bibinfo{journal}{Phys. Rev. Lett.} \textbf{\bibinfo{volume}{{\bf 73}}},
  \bibinfo{pages}{1376} (\bibinfo{year}{1994});
  \bibinfo{journal}{Phys. Rev. {\rm E}} \textbf{\bibinfo{volume}{{\bf 51}}},
  \bibinfo{pages}{4626} (\bibinfo{year}{1995}{\natexlab{a}});
  \bibinfo{journal}{{\it ibid.}} \textbf{\bibinfo{volume}{{\bf 52}}},
  \bibinfo{pages}{4134} (\bibinfo{year}{1995}{\natexlab{b}}).

\bibitem[{\citenamefont{Foffi et~al.}((2004))\citenamefont{Foffi, G{\"{o}}tze,
  Sciortino, Tartaglia, and Voigtmann}}]{Foffi2004}
\bibinfo{author}{\bibfnamefont{G.}~\bibnamefont{Foffi}},
  \bibinfo{author}{\bibfnamefont{W.}~\bibnamefont{G{\"{o}}tze}},
  \bibinfo{author}{\bibfnamefont{F.}~\bibnamefont{Sciortino}},
  \bibinfo{author}{\bibfnamefont{P.}~\bibnamefont{Tartaglia}},
  \bibnamefont{and}
  \bibinfo{author}{\bibfnamefont{T.}~\bibnamefont{Voigtmann}},
  \bibinfo{journal}{Phys. Rev. {\rm E}} \textbf{\bibinfo{volume}{{\bf 69}}},
  \bibinfo{pages}{011505} (\bibinfo{year}{2004}).

\bibitem[{\citenamefont{Flenner and
  Szamel}((2005){\natexlab{a}})}]{Flenner2005d}
\bibinfo{author}{\bibfnamefont{E.}~\bibnamefont{Flenner}} \bibnamefont{and}
  \bibinfo{author}{\bibfnamefont{G.}~\bibnamefont{Szamel}},
  \bibinfo{journal}{Phys. Rev. {\rm E}} \textbf{\bibinfo{volume}{{\bf 72}}},
  \bibinfo{pages}{031508} (\bibinfo{year}{2005}{\natexlab{a}}).

\bibitem[{\citenamefont{Kob et~al.}((2002))\citenamefont{Kob, Nauroth, and
  Sciortino}}]{Kob2002}
\bibinfo{author}{\bibfnamefont{W.}~\bibnamefont{Kob}},
  \bibinfo{author}{\bibfnamefont{M.}~\bibnamefont{Nauroth}}, \bibnamefont{and}
  \bibinfo{author}{\bibfnamefont{F.}~\bibnamefont{Sciortino}},
  \bibinfo{journal}{J. Non-Cryst. Solids} \textbf{\bibinfo{volume}{{\bf
  307-310}}}, \bibinfo{pages}{181} (\bibinfo{year}{2002}),
  \bibinfo{note}{{}}.

\bibitem[{\citenamefont{Flenner and
  Szamel}((2005){\natexlab{b}})}]{Flenner2005e}
\bibinfo{author}{\bibfnamefont{E.}~\bibnamefont{Flenner}} \bibnamefont{and}
  \bibinfo{author}{\bibfnamefont{G.}~\bibnamefont{Szamel}},
  \bibinfo{journal}{Phys. Rev. {\rm E}} \textbf{\bibinfo{volume}{{\bf 72}}},
  \bibinfo{pages}{011205} (\bibinfo{year}{2005}{\natexlab{b}}).

\bibitem[{\citenamefont{Sastry et~al.}((1998))\citenamefont{Sastry,
  Debenedetti, and Stillinger}}]{sastry1998}
\bibinfo{author}{\bibfnamefont{S.}~\bibnamefont{Sastry}},
  \bibinfo{author}{\bibfnamefont{P.~G.} \bibnamefont{Debenedetti}},
  \bibnamefont{and} \bibinfo{author}{\bibfnamefont{F.~H.}
  \bibnamefont{Stillinger}}, \bibinfo{journal}{Nature}
  \textbf{\bibinfo{volume}{{\bf 393}}}, \bibinfo{pages}{554}
  (\bibinfo{year}{1998}).

\bibitem[{\citenamefont{Brumer and Reichman}((2004))}]{Brumer2004b}
\bibinfo{author}{\bibfnamefont{Y.}~\bibnamefont{Brumer}} \bibnamefont{and}
  \bibinfo{author}{\bibfnamefont{D.~R.} \bibnamefont{Reichman}},
  \bibinfo{journal}{Phys. Rev. {\rm E}} \textbf{\bibinfo{volume}{{\bf 69}}},
  \bibinfo{pages}{041202} (\bibinfo{year}{2004}).

\bibitem[{\citenamefont{Mayer et~al.}((2006))\citenamefont{Mayer, Miyazaki, and
  Reichman}}]{Mayer2006b}
\bibinfo{author}{\bibfnamefont{P.}~\bibnamefont{Mayer}},
  \bibinfo{author}{\bibfnamefont{K.}~\bibnamefont{Miyazaki}}, \bibnamefont{and}
  \bibinfo{author}{\bibfnamefont{D.~R.} \bibnamefont{Reichman}},
  \bibinfo{journal}{Phys. Rev. Lett.} \textbf{\bibinfo{volume}{{\bf 97}}},
  \bibinfo{pages}{095702} (\bibinfo{year}{2006}).

\bibitem[{\citenamefont{Bhattacharyya
  et~al.}((2008))\citenamefont{Bhattacharyya, Bagchi, and
  Wolynes}}]{Bhattacharyya2008}
\bibinfo{author}{\bibfnamefont{S.~M.} \bibnamefont{Bhattacharyya}},
  \bibinfo{author}{\bibfnamefont{B.}~\bibnamefont{Bagchi}}, \bibnamefont{and}
  \bibinfo{author}{\bibfnamefont{P.~G.} \bibnamefont{Wolynes}},
  \bibinfo{journal}{Proc. Natl. Acad. Sci. U. S. A.}
  \textbf{\bibinfo{volume}{{\bf 105}}}, \bibinfo{pages}{16077}
  (\bibinfo{year}{2008}).

\bibitem[{\citenamefont{Kumar et~al.}((2006))\citenamefont{Kumar, Szamel, and
  Douglas}}]{Kumar2006}
\bibinfo{author}{\bibfnamefont{S.~K.} \bibnamefont{Kumar}},
  \bibinfo{author}{\bibfnamefont{G.}~\bibnamefont{Szamel}}, \bibnamefont{and}
  \bibinfo{author}{\bibfnamefont{J.~F.} \bibnamefont{Douglas}},
  \bibinfo{journal}{J. Chem. Phys.} \textbf{\bibinfo{volume}{{\bf 124}}},
  \bibinfo{pages}{214501} (\bibinfo{year}{2006}).

\bibitem[{\citenamefont{Voigtmann et~al.}((2004))\citenamefont{Voigtmann,
  Puertas, and Fuchs}}]{Voigtmann2004}
\bibinfo{author}{\bibfnamefont{T.}~\bibnamefont{Voigtmann}},
  \bibinfo{author}{\bibfnamefont{A.~M.} \bibnamefont{Puertas}},
  \bibnamefont{and} \bibinfo{author}{\bibfnamefont{M.}~\bibnamefont{Fuchs}},
  \bibinfo{journal}{Phys. Rev. {\rm E}} \textbf{\bibinfo{volume}{{\bf 70}}},
  \bibinfo{pages}{061506} (\bibinfo{year}{2004}).

\bibitem[{\citenamefont{Gleim et~al.}((1998))\citenamefont{Gleim, Kob, and
  Binder}}]{Gleim1998}
\bibinfo{author}{\bibfnamefont{T.}~\bibnamefont{Gleim}},
  \bibinfo{author}{\bibfnamefont{W.}~\bibnamefont{Kob}}, \bibnamefont{and}
  \bibinfo{author}{\bibfnamefont{K.}~\bibnamefont{Binder}},
  \bibinfo{journal}{Phys. Rev. Lett.} \textbf{\bibinfo{volume}{{\bf 81}}},
  \bibinfo{pages}{4404} (\bibinfo{year}{1998}).

\bibitem[{\citenamefont{Shiroiwa et~al.}(())\citenamefont{Shiroiwa, Ikeda, and
  Miyazaki}}]{Shiroiwa2010}
\bibinfo{author}{\bibfnamefont{S.}~\bibnamefont{Shiroiwa}},
  \bibinfo{author}{\bibfnamefont{A.}~\bibnamefont{Ikeda}}, \bibnamefont{and}
  \bibinfo{author}{\bibfnamefont{K.}~\bibnamefont{Miyazaki}},
  \bibinfo{journal}{unpublished} \textbf{\bibinfo{volume}{{\bf }}}.

\bibitem[{\citenamefont{Torquato and Stillinger}((2003))}]{Torquato2003}
\bibinfo{author}{\bibfnamefont{S.}~\bibnamefont{Torquato}} \bibnamefont{and}
  \bibinfo{author}{\bibfnamefont{F.~H.} \bibnamefont{Stillinger}},
  \bibinfo{journal}{Phys. Rev. {\rm E}} \textbf{\bibinfo{volume}{{\bf 68}}},
  \bibinfo{pages}{041113} (\bibinfo{year}{2003});
\bibinfo{author}{\bibfnamefont{A.}~\bibnamefont{Donev}},
  \bibinfo{author}{\bibfnamefont{F.~H.} \bibnamefont{Stillinger}},
  \bibnamefont{and} \bibinfo{author}{\bibfnamefont{S.}~\bibnamefont{Torquato}},
  \bibinfo{journal}{Phys. Rev. Lett.} \textbf{\bibinfo{volume}{{\bf 95}}},
  \bibinfo{pages}{090604} (\bibinfo{year}{2005}).

\bibitem[{\citenamefont{Ediger}((2000))}]{ediger2000}
\bibinfo{author}{\bibfnamefont{M.~D.} \bibnamefont{Ediger}},
  \bibinfo{journal}{Annu. Rev. Phys. Chem.} \textbf{\bibinfo{volume}{{\bf
  51}}}, \bibinfo{pages}{99} (\bibinfo{year}{2000}).

\bibitem[{\citenamefont{Hansen and McDonald}((1986))}]{hansen1986}
\bibinfo{author}{\bibfnamefont{J.~P.} \bibnamefont{Hansen}} \bibnamefont{and}
  \bibinfo{author}{\bibfnamefont{I.~R.} \bibnamefont{McDonald}},
  \emph{\bibinfo{title}{{"Theory of simple liquids"}}}
  (\bibinfo{publisher}{Academic Press}, \bibinfo{year}{1986}).

\bibitem[{\citenamefont{Cates et~al.}((2004))\citenamefont{Cates, Fuchs, Kroy,
  Poon, and Puertas}}]{Cates2004}
\bibinfo{author}{\bibfnamefont{M.~E.} \bibnamefont{Cates}},
  \bibinfo{author}{\bibfnamefont{M.}~\bibnamefont{Fuchs}},
  \bibinfo{author}{\bibfnamefont{K.}~\bibnamefont{Kroy}},
  \bibinfo{author}{\bibfnamefont{W.~C.~K.} \bibnamefont{Poon}},
  \bibnamefont{and} \bibinfo{author}{\bibfnamefont{A.~M.}
  \bibnamefont{Puertas}}, \bibinfo{journal}{J. Phys.: Condens. Matter}
  \textbf{\bibinfo{volume}{{\bf 16}}}, \bibinfo{pages}{S4861}
  (\bibinfo{year}{2004}).

\bibitem[{\citenamefont{Reichman and Charbonneau}((2005))}]{Reichman2005}
\bibinfo{author}{\bibfnamefont{D.~R.} \bibnamefont{Reichman}} \bibnamefont{and}
  \bibinfo{author}{\bibfnamefont{P.}~\bibnamefont{Charbonneau}},
  \bibinfo{journal}{J. Stat. Mech.}
  \bibinfo{pages}{P05013} (\bibinfo{year}{2005}).

\end{thebibliography}
\end{document}